\documentclass[%
 preprint,
 amsmath,amssymb,
 aps,
prc,
]{aastex63}
\usepackage{bm}% bold math
\usepackage{mathrsfs}
\usepackage{natbib}
\usepackage{graphicx,epsf,amssymb,amsbsy,amsfonts,amssymb,amsmath}
\usepackage{graphicx,epsf,amssymb,bbm,amsbsy,amsfonts,amssymb,amsmath}
\usepackage{slashed}
\usepackage{color,soul}
\usepackage[colorinlistoftodos]{todonotes}
\usepackage{comment}

\usepackage[british]{datetime2}

\newcommand{\be}{\begin{equation}}
\newcommand{\ee}{\end{equation}}
\newcommand\beq{\begin{eqnarray}}
\newcommand\eeq{\end{eqnarray}}

\newcommand{\mybar}[1]

\begin{document}
%\preprint{INT-PUB-16-042}
\title{Mass and radius relations of quarkyonic stars using an excluded volume model}% Force line breaks with \\
%\thanks{A footnote to the article title}%
%\author{David B. Kaplan}
%\email{dbkaplan@uw.edu}
%\affiliation{%
%Institute for Nuclear Theory, University of Washington, Seattle, WA 98195  
%}%
%\author{Sanjay Reddy}
%\email{sareddy@uw.edu}
%\affiliation{
%Institute for Nuclear Theory, University of Washington, Seattle, WA 98195  
%}%
\author{Srimoyee Sen}%
\email{srimoyee08@gmail.com}
\affiliation{
Department of Physics and Astronomy, Iowa State University, Ames, IA  50010
}%

\author{Lars Sivertsen}%
\email{lars@iastate.edu}
\affiliation{
Department of Physics and Astronomy, Iowa State University, Ames, IA  50010
}
%\date{\today}
%\date{\today}% It is always \today, today,
             %  but any date may be explicitly specified

\begin{abstract}

%An `excluded volume' dynamical model was recently proposed to build a quarkyonic description for isospin symmetric dense matter where the quasiparticle degrees of freedom are two flavors of nucleons and two flavors of quarks. 
 Inspired by the excluded volume model for isospin symmetric quarkyonic matter \cite{PhysRevC.101.035201}, we construct an `excluded volume' model for a charge neutral quarkyonic phase whose hadronic sector contains only neutrons. We refer to this model as quarkyonic neutron matter. We compute the equation of state for this model and solve the Tolman-Oppenhermer-Volkoff equations to obtain mass and radius relations relevant for neutron stars. The most straightforward extension of the model for symmetric quarkyonic matter \cite{PhysRevC.101.035201} to quarkyonic neutron matter does not satisfy the mass radius constraints from neutron star measurements.
However, we show that by incorporating appropriate nuclear interactions in the excluded volume model one can produce mass-radius relations that lie within the constraints obtained from gravitational waves of binary neutron star mergers and maximum mass measurements of neutron stars. 
 
\end{abstract}

%\maketitle

\section{Introduction}
Neutron star observations \cite{Watts:2016uzu, Ozel:2016oaf, Abbott:2018exr, Demorest:2010bx, Antoniadis:2013pzd, Annala:2017llu,De:2018uhw, Tews:2018iwm, Raaijmakers:2019dks} suggest that the equation of state of dense matter is soft at low baryon density (lower than a few times the nuclear saturation density) and is stiff at higher baryon densities \cite{DrischlerChristian2020Lmar}. This behavior in turn dictates the behavior of speed of sound as a function of baryon density. At very low density the speed of sound is much smaller than $1$ whereas at asymptotically high densities, in weakly coupled quark matter the speed of sound squared approaches the conformal bound $1/3$ from below with increasing baryon density. The soft-stiff nature of the equation of state implies that the speed of sound is not a monotonically increasing function of baryon density. Instead it is likely to have a peak at some intermediate density where the speed of sound squared will be larger than the conformal bound of $1/3$ \cite{Bedaque:2014sqa}. In general it is difficult to model this behavior of the equation of state using purely hadronic matter \cite{Akmal:1998cf, Hebeler:2009iv, Gandolfi:2011xu}. This implies that the hadronic description of dense matter breaks down at densities that are well within the reach of neutron stars. Such a conclusion is also supported by insights from chiral effective field theory
which estimates the breakdown density of a purely nucleonic description to be two times the nuclear saturation density \cite{McLerran:2018hbz}.
Moreover, even when quark degrees of freedom are taken into account at higher density, it may still be difficult to model the soft-stiff nature of the equation of state if quark onset is accompanied by a phase transition \cite{Han:2018mtj, Alford:2004pf, Chatziioannou:2019yko}.

%\footnotetext[1]{A quark-hadron crossover model constructed in \cite{Baym:2019iky} was also found to agree well with neutron star mass radius constraints}

Recently a model of dense QCD, known as quarkyonic matter \cite{McLerran:2007qj, Hidaka:2008yy} was found to explain the mass and radius measurements of neutron stars rather naturally \cite{McLerran:2018hbz}.\footnote{A quark-hadron crossover model constructed in \cite{Baym:2019iky} was also found to agree well with neutron star mass radius constraints} This makes quarkyonic matter a particularly attractive candidate for modeling the behavior of cold dense QCD. At low baryon density quarkyonic matter resembles nuclear matter. However, at high density the fermi distribution function of quarkyonic matter is what sets it apart from purely hadronic or quark matter. Within the quarkyonic matter framework, nucleon and quark degrees of freedom are described with a single fermi distribution function. More specifically, the fermi distribution constitutes of two parts: an inner sphere of quark quasiparticles and an outer shell of hadronic quasiparticles. This fermi distribution is motivated by the idea that even at high density, confinement persists for the low energy degrees of freedom near the fermi surface. As a result one expects to find quark degrees of freedom deep inside the fermi sea while the degrees of freedom near the fermi surface remain hadronic. 

\begin{figure}
	\centering
	\includegraphics[width=0.7\textwidth]{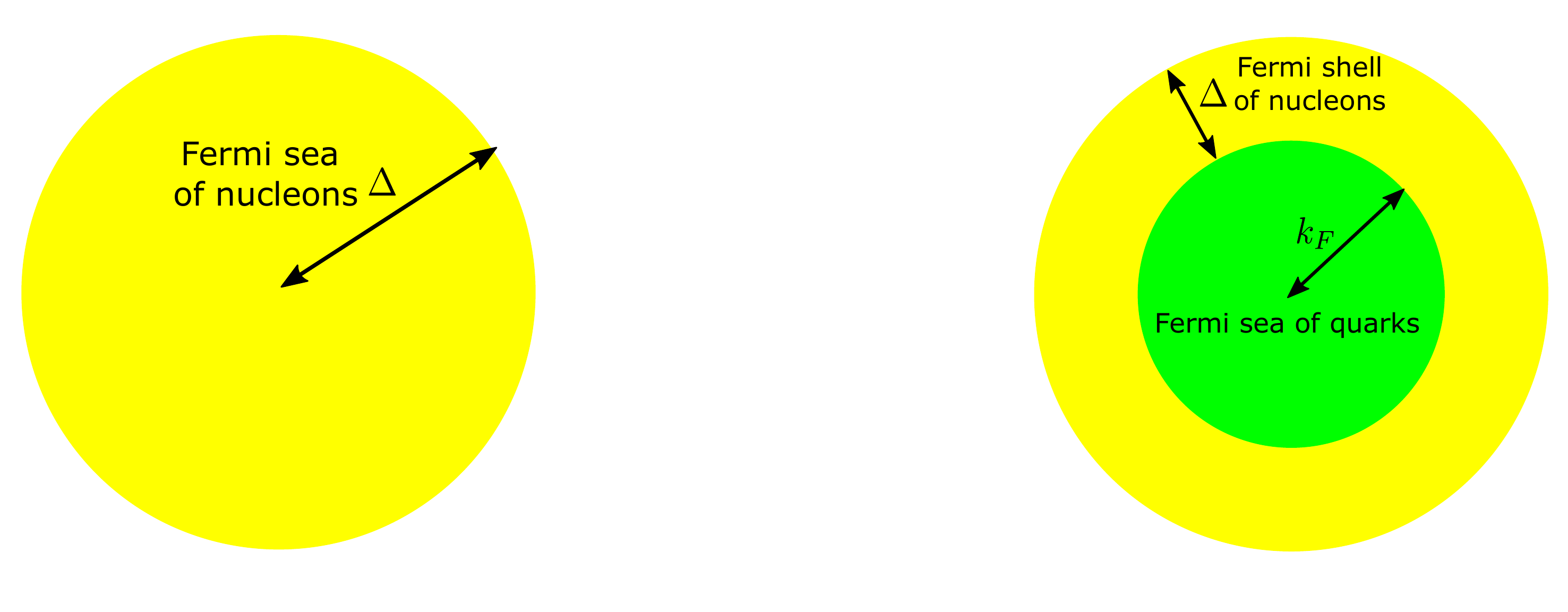}
	\caption{Illustration of the fermi sphere for quarkyonic matter at low baryon density before the quark onset (left panel) and at high density after quark onset (right panel).}
	\label{shell}
\end{figure}

A dynamical model for quarkyonic matter was developed in \cite{PhysRevC.101.035201} where both nucleons and quarks appear as quasiparticles. The model in \cite{PhysRevC.101.035201} is isospin symmetric where the hadronic sector corresponds to symmetric nuclear matter. In this model a  particular configuration of the quarkyonic fermi distribution which is described by the radius of the inner quark fermi sphere and the shell width of the nucleon states, is tied to a particular density of quarks and nucleons. In other words, the radius of the quark fermi sphere and the nucleon shell width in a particular quarkyonic configuration fix the quark density and the nucleon density for that configuration. The sum of the quark and the nucleon density then gives the total baryon density. The model of \cite{PhysRevC.101.035201} also proposes an energy density functional which can be expressed as a function of either the quark density or the nucleon density for a fixed baryon density. This energy density functional treats nucleons as hard spheres of size $v_0$ and quarks as non-interacting point particles. Minimizing the energy density functional with respect to either the quark density or the nucleon density amounts to minimizing the energy density functional with respect to various configurations of the quarkyonic fermi distribution. The minimization procedure produces the optimal configuration for the quarkyonic fermi sphere for a fixed baryon density that results in the lowest energy configuration. At low baryon density we expect the lowest energy configuration to be purely hadronic which corresponds to a fermi sphere of hadrons. As the baryon density rises, a finite density of quarks appears giving rise to a shell of hadrons in the fermi distribution. The left panel of Figure \ref{shell} shows the purely hadronic fermi distribution at low density whereas the fermi distribution in the right panel of Figure \ref{shell} exhibits a finite density of quarks which corresponds to a higher baryon density.   

The dynamical model of \cite{PhysRevC.101.035201} was found to exhibit the desired behavior in the speed of sound i.e. the speed is small at very small densities and then goes through a peak at intermediate densities to eventually approach the conformal bound from below at very high density. Note that the speed of sound is one of the many ingredients that determine the mass and radius of neutron stars. In particular one needs to know the energy density or the pressure as a function of the total baryon density to produce mass and radius relations for neutron stars. In this paper we examine whether the dynamical model proposed in \cite{PhysRevC.101.035201} or some variation of it can give rise to mass and radius of neutron stars that fall within observational constraints. It is of course easy to see that the model in \cite{PhysRevC.101.035201} in its current form will not produce the correct behavior for mass and radius of neutron stars. One of the prime reasons behind this is that the model in \cite{PhysRevC.101.035201} is isospin symmetric. We will need to impose charge neutrality as well as consider a neutron rich hadronic sector in order to produce mass radius relations that resemble that of a neutron star. In this paper we take the first step towards achieving this goal. We develop a model for quarkyonic matter where the degrees of freedom in the hadronic sector includes only neutrons. In the quark sector we have up and down quarks and their densities are constrained by charge neutrality. We will not consider beta equilibrium in this paper and reserve it for future work. Following \cite{McLerran:2018hbz} we will treat the quarks as free particles in this work. We plan to include interactions between the quarks and the nucleons in future work. We do however include nuclear interactions while constructing the excluded volume models of quarkyonic neutron matter discussed in this paper. In fact, the inclusion of appropriate attractive and repulsive interactions consistent with microscopic calculations of neutron matter \cite{PhysRevC.85.032801} is crucial in order to obtain quarkyonic neutron matter models that satisfy neutron star constraints.

The organization of the paper is as follows. We begin with a discussion on some conceptual points involved in constructing the EOS of quarkyonic matter. In the second section we will review the dynamical model in \cite{PhysRevC.101.035201} for isospin symmetric quarkyonic matter. Then we will construct an excluded volume model for charge neutral quarkyonic matter with a pure neutron hadronic sector. As stated earlier the simplest extension of the model in \cite{PhysRevC.101.035201} incorporating charge neutrality that restricts the hadronic degrees of freedom to neutrons and constrains the up and down quark densities does not ensure that the model satisfies NS mass radius constraints. In the next section we discuss further modifications to the excluded volume model that involve incorporating neutron matter interactions to obtain mass radius relations that satisfy NS constraints. We end with a conclusion and summary of future work.

%Although one in the recent years have a fairly good understanding of nuclear matter at low densities, there is still missing a complete model for dense matter beyond saturation. Recent discoveries of neutron star mergers using gravitational waves have provided new and narrower constraints on their radii, as well as the pressure at 3 times the saturation density inside neutron stars \cite{Stringent_constraints}. Previous measurements of neutron star masses around and above 2 solar masses \cite{Cromartie2020}\cite{Antoniadis1233232}\cite{Demorest2010} is making it less probable that the equation of state is purely hadronic (citation needed). A recent discovery even opens the possibility of neutron stars with masses as high as 2.5 solar masses \cite{Abbott:2020khf}. This, in addition to arguments regarding the large $N_\text{c}$ limit of QCD \cite{MCLERRAN200783}, has motivated  models of quarkyonic matter, such as \cite{PhysRevC.101.035201} and \cite{PhysRevLett.122.122701}.
\section{Discussion of conceptual points}
Incorporating neutron matter interactions in a dynamical model for quarkyonic matter offers numerous challenges. Before we begin discussing the details of this dynamical model, it is important to underline these challenges and associated conceptual points which will be crucial to understanding the importance of the construction presented in this paper. 
\begin{itemize}
%\item \textbf{ORIGINAL BULLET POINT: } One of these challenges of incorporating nuclear or neutron matter interactions in a dynamical model for quarkyonic matter relates to obtaining a single equation of state that correctly interpolates between quarkyonic behavior at high density and a low density nuclear equation of state (EOS) consistent with neutron matter interactions. To achieve this interpolation, it may be tempting to apply the Maxwell construction between the nuclear EOS at low density and the quarkyonic EOS at high density. However, such a construction is not well motivated in light of the fact that the dynamical model of quarkyonic matter constructed in \cite{PhysRevC.101.035201} does not exhibit any phase transition at the quark onset by design. More specifically, the energy density as a function of the baryon density in the dynamical model is convex eliminating the possibility of a phase transition. As a result, resorting to the Maxwell construction is not the ideal way to incorporate nuclear interactions in the dynamical model for quarkyonic matter. In this paper, we devise an alternative strategy to incorporate neutron matter interactions in the quarkyonic EOS while avoiding the Maxwell construction. %This EOS produces the correct low density behavior consistent with nuclear interactions while avoiding Maxwell construction. 

\item One of these challenges of incorporating nuclear or neutron matter interactions in a dynamical model for quarkyonic matter relates to obtaining a single equation of state that correctly interpolates between quarkyonic behavior at high density and a low density nuclear equation of state (EOS) consistent with neutron matter interactions. While constructing such interpolations, it is not uncommon for models to exhibit a phase transition between nuclear matter and quark matter, where the EOS in the crossover region is found using the Maxwell construction. This approach was adopted for excluded volume quarkyonic matter in \cite{Duarte:2020kvi}. However, such a construction is not well motivated in light of the fact that the dynamical model of quarkyonic matter constructed in \cite{PhysRevC.101.035201} does not exhibit any phase transition at the quark onset by design. More specifically, the energy density as a function of the baryon density in the dynamical model is convex eliminating the possibility of a phase transition. As a result, we in this paper devise an alternative strategy to incorporate neutron matter interactions in the quarkyonic EOS while avoiding the Maxwell construction.

\item There is a second challenge to incorporating nuclear interactions in the excluded volume model for quarkyonic matter. This relates to the fact that although the sole purpose for introducing an excluded volume potential for the nucleons in \cite{PhysRevC.101.035201} was to disfavor nucleons only at high density thereby facilitating the appearance of quarks at those densities, the excluded volume potential ends up significantly altering the low density nuclear EOS. As a result the EOS obtained from the dynamical excluded volume model in \cite{PhysRevC.101.035201} is not consistent with the phenomenology of neutron matter interactions at low density. Designing a dynamical model for quarkyonic matter which produces a low density EOS consistent with nuclear interactions, thus involves, curtailing the effect of the excluded volume potential on the low density EOS while retaining its impact on the high density part of the EOS which enables the onset of quarks. In the sections that follow we elaborate in detail how this is achieved in the models we have constructed. 

\item Finally, it was observed in the dynamical model of \cite{PhysRevC.101.035201} that the simplest version of the excluded volume potential produces an abrupt onset of quarks which is undesired as it produces unphysical behavior for the speed of sound. Note that, it is incorrect to attribute this unphysical behavior for the speed of sound to the existence of a phase transition since the quarkyonic model is constructed precisely to avoid the scenario of a phase transition between nuclear matter and quark matter. An attempt was made in \cite{PhysRevC.101.035201} to cure the unphysical behavior of the speed of sound at quark onset by introducing a regulator for the quark density of states which would result in a gradual onset of quarks. Even though such a regulator eliminates the unphysical behavior exhibited by the speed of sound at quark onset, it introduces another undesirable feature where quarks are produced for baryon density smaller than the saturation density. As described in this paper, we modify the regulator introduced in \cite{PhysRevC.101.035201} to avoid quark production at low density while also retaining the gradual onset of quarks beyond a critical baryon density
\end{itemize}

\section{Dynamical model for symmetric quarkyonic matter}
In this section we first describe the dynamical model of \cite{PhysRevC.101.035201} which we refer to as quarkyonic symmetric matter (QSM) and then move on to discuss the limitations of this model. In the latter part of this section we introduce modifications to this model that rectify some of the undesired features of \cite{PhysRevC.101.035201}. 
\subsection{Review of the QSM model}
In this model, neutrons, protons, up quarks and down quarks appear as quasiparticles. Both flavors of nucleons and quarks are considered to be degenerate and charge neutral. There are two main ingredients of the model for QSM : 
\begin{enumerate}
\item the quarkyonic phase should arise dynamically as a result of a minimization procedure applied to an energy density functional.
\item the energy density functional models the nucleons as hard spheres. 
\end{enumerate}
In this model the quarkyonic fermi distribution is imposed by construction, i.e. when there are quarks present, they are assumed to occupy an inner spherical volume inside the fermi sphere and are surrounded by a shell of nucleon states. However, the information of whether there are any quarks present at all at a particular baryon density is what is determined dynamically in this model. The density of quarks is related to the radius of the inner fermi sphere assuming the quarks are noninteracting. Similarly, the density of nucleons is related to the shell of nucleons outside the quark fermi sphere. The total baryon density is the sum of the quark and the nucleon densities. Hence, for a fixed baryon density one can vary the radius of the inner quark sphere thus varying the proportion of quarks while keeping the total baryon density fixed. The model also includes an energy density functional in terms of quark and nucleon densities. This energy density functional is then minimized with respect to the quark and nucleon densities for a fixed total baryon density. This constrained minimization results in the lowest energy configuration for the quarkyonic fermi distribution. 
In order to model quarkyonic behavior adequately we need an energy functional that produces purely hadronic matter at low baryon density and realizes a finite density of quarks only when a certain critical baryon density is reached. In the dynamical model for QSM, this was achieved by describing nucleons as hard spheres, i.e. if each nucleon is a hard sphere with volume $v_0$, then there is an upper bound on the number of nucleons that can be fit in a box of volume $V$. This sets the maximum critical density achievable by hadronic degrees of freedom which give way to quarks in the system once the total baryon density exceeds this critical density. In practice the hard core interaction of the nucleons is encoded in the inner and outer radii of the fermi shell that the nucleons occupy. This is achieved as follows. At first we define an excluded density for the nucleons given by
\beq
n_\text{ex}^\text{N} = \frac{n_\text{B}^\text{N}}{1-n_\text{B}^\text{N}v_0} \equiv \frac{n_\text{B}^\text{N}}{1-\frac{n_\text{B}^\text{N}}{n_0}}.
\label{nex}
\eeq
Here $n_0\equiv1/v_0$ is the hardcore density and $n_B^N$ is the baryon density in nucleons. We will take the hardcore density to be larger than the saturation density which we denote as $\rho_0$.
The excluded density keeps account of the fact that due to the finite size of the nucleons, in a box of volume $V$ with $N$ nucleons, the $(N+1)^{\text{th}}$ nucleon can only occupy a volume of $V-N v_0$. The result of the finite size of the nucleons is to raise the effective density of the nucleons. We then express the nucleon fermi momenta in terms of the excluded density using 
\beq
n_{\text{ex}}^{\text{N}}=4\int_{k_F}^{k_F+\Delta} \frac{d^3k}{(2\pi)^3},
\label{fermi1}
\eeq
where $k_F$ is the bottom of the nucleon fermi shell and $\Delta$ is the width of the shell such that $k_F+\Delta$ is the top of the nucleon fermi surface. Note that the factor of $4$ accounts for the spin and two flavors of nucleons. We can then relate the bottom of the nucleon fermi shell $k_F$ and the shell width $\Delta$ as
\beq
\Delta = \bigg(\frac{3\pi^2}{2}n_\text{ex}^\text{N}+k_\text{F}^3\bigg)^{\frac{1}{3}}-k_\text{F}.
\label{delta}
\eeq
This is how the hardcore interaction of the nucleons is captured in the inner and outer radii of the nucleon shell. 
The energy density of the nucleons in the excluded volume model is then given by
\beq
\epsilon_N = 4\bigg(1-\frac{n_\text{B}^\text{N}}{n_0}\bigg)\int_{k_\text{F}}^{k_\text{F}+\Delta}\frac{\text{d}^3k}{(2\pi)^3}\sqrt{M_\text{N}^2+k^2},
%\label{e1}
\eeq
where $M_N$ is the mass of a nucleon. The factor of $1-n_B^N/n_0$ in the expression for energy density takes into account of the fact that $N$ nucleons have an available volume of $V-N v_0$ in a box of volume $V$. 
Treating the quarks as noninteracting point particles of mass $m$, the baryon density stored in quarks can be written as 
\beq
n_B^Q&=&4\int_0^\frac{k_F}{N_c}\frac{d^3k}{(2\pi)^3}\nonumber\\
&=&\frac{2}{3\pi^2}\left(\frac{k_F}{N_c}\right)^3.
\label{nq1}
\eeq
The degeneracy factor of $4$ in the first line of Eq. \ref{nq1} corresponds to a factor of $2$ coming from the two spin degrees of freedom and $2$ flavors of quarks.
The mass of the quarks $m$ is related to the mass of the nucleons by $M_N=m N_c$ where $N_c$ is the number of colors and we take $N_c=3$ in the rest of the analysis. The total baryon density is given by 
\beq
n_B=n_B^Q+n_B^N.
\label{nb}
\eeq
The energy density functional is the sum of the energy densities of the nucleons and quarks for a particular baryon density and is given by
\beq
\epsilon_{\text{QSM}}= 4\bigg(1-\frac{n_\text{B}^\text{N}}{n_0}\bigg)\int_{k_\text{F}}^{k_\text{F}+\Delta}\frac{\text{d}^3k}{(2\pi)^3}\sqrt{M_\text{N}^2+k^2}+4N_\text{C}\int_0^{k_\text{F}/N_\text{C}}\frac{d^3k}{(2\pi)^3}\sqrt{m^2+k^2},
\label{e2}
\eeq
where the subscript QSM stands for quarkyonic symmetric matter.
Note that using Eq. \ref{nex} and \ref{delta} one can express $\Delta$ as a function of the total baryon density and $k_F$ or $n_B^Q$. As a result the energy functional in Eq. \ref{e2} can be expressed solely as a function of $k_F$ ($n_B^Q$) as well. One can then minimize the energy functional $\epsilon_{\text{QSM}}$ with respect to $k_F$ or the quark density for a fixed baryon density to arrive at the equilibrium quark density. 

The energy density functional in Eq. \ref{e2} gives rise to purely nucleonic configurations for $n_B <n_0$ and quarkyonic configurations for $n_B>n_0$. A simple way to see this is to consider purely hadronic and purely quark configurations for the fermi sphere and compare their energy densities. For an all nucleon configuration $n_B^Q=0$ the energy density in the non-relativistic limit is given by 
\beq
\epsilon_N=\frac{1}{5\pi^2 M_N}\left(\frac{3\pi^2 n_B}{2}\right)^{5/3}\frac{1}{\left(1-\frac{n_B}{n_0}\right)^{2/3}}.
\label{en1}
\eeq
Similarly an all quark configuration in the same limit has $n_B^N=0$ and the energy density is given by
\beq
\epsilon_Q=\frac{N_c^2}{5\pi^2M_N}\left(\frac{3\pi^2 n_B}{2}\right)^{5/3}.
\label{en2}
\eeq
Comparing Eq. \ref{en1} and \ref{en2} one can see that for $n_B \ll n_0$
the energy cost of storing baryon density in nucleon degrees of freedom is suppressed by a factor of $N_c^2$ compared to the energy cost of storing them in quark degrees of freedom. Thus at low baryon density the minimizing configuration is where baryon density is stored in nucleon degrees of freedom. However, as $n_B$ approaches $n_0$, the singularity in $\epsilon_N$ increases the energy cost of storing baryon density in nucleon degrees of freedom. When $(1-\frac{n_B}{n_0})^{-2/3}\sim N_c^2$, it becomes favorable to store some of the baryon density in quarks and the purely hadronic fermi distribution gives way to quarkyonic fermi distribution. 
\iffalse
We also note that for either pure neutron- or pure symmetric nuclear matter, i.e. no quarks present, we can rewrite the prefactor $(3\pi^2/2)^{5/3}\rightarrow (6\pi^2/g)^{5/3}$, where $g=2$ for neutron matter and $g=4$ for symmetric nuclear matter. This gives us the pressure
\beq
P_N^{(g)} = n_B\frac{\text{d}\epsilon_N}{\text{d}n_B}-\epsilon_N = \frac{2}{15\pi^2 M_N}\bigg(\frac{3\pi^2 n_B}{g}\bigg)^{5/3}\frac{1}{\left(1-\frac{n_B}{n_0}\right)^{5/3}},
\eeq
\hl{which shows us that for a given, small, baryon density, the pressure of symmetric nuclear matter is smaller than the pressure from neutron matter. We will later see that this manifest it self in giving neutron stars made of symmetric nuclear matter a smaller radius than a pure neutron star.}
\fi
\subsubsection{Speed of sound and regulator}
\begin{figure}
	\includegraphics[width=\textwidth]{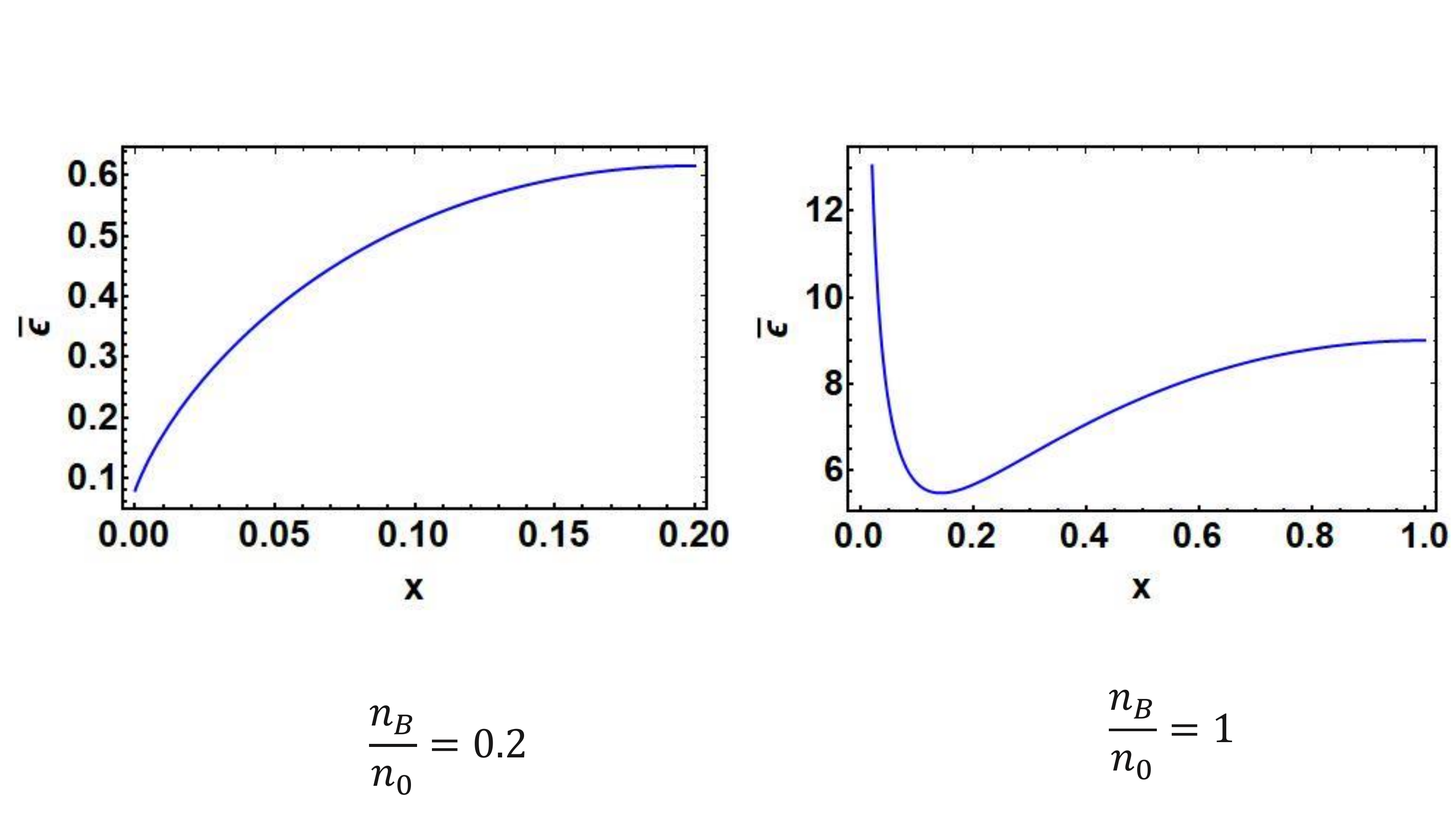}
	\caption{The energy density functional of Eq. \ref{xye} as a function of the normalized quark density $n_B^Q/n_0=x$.}
	\label{min1}
\end{figure}

\begin{figure}
	\includegraphics[width=\textwidth]{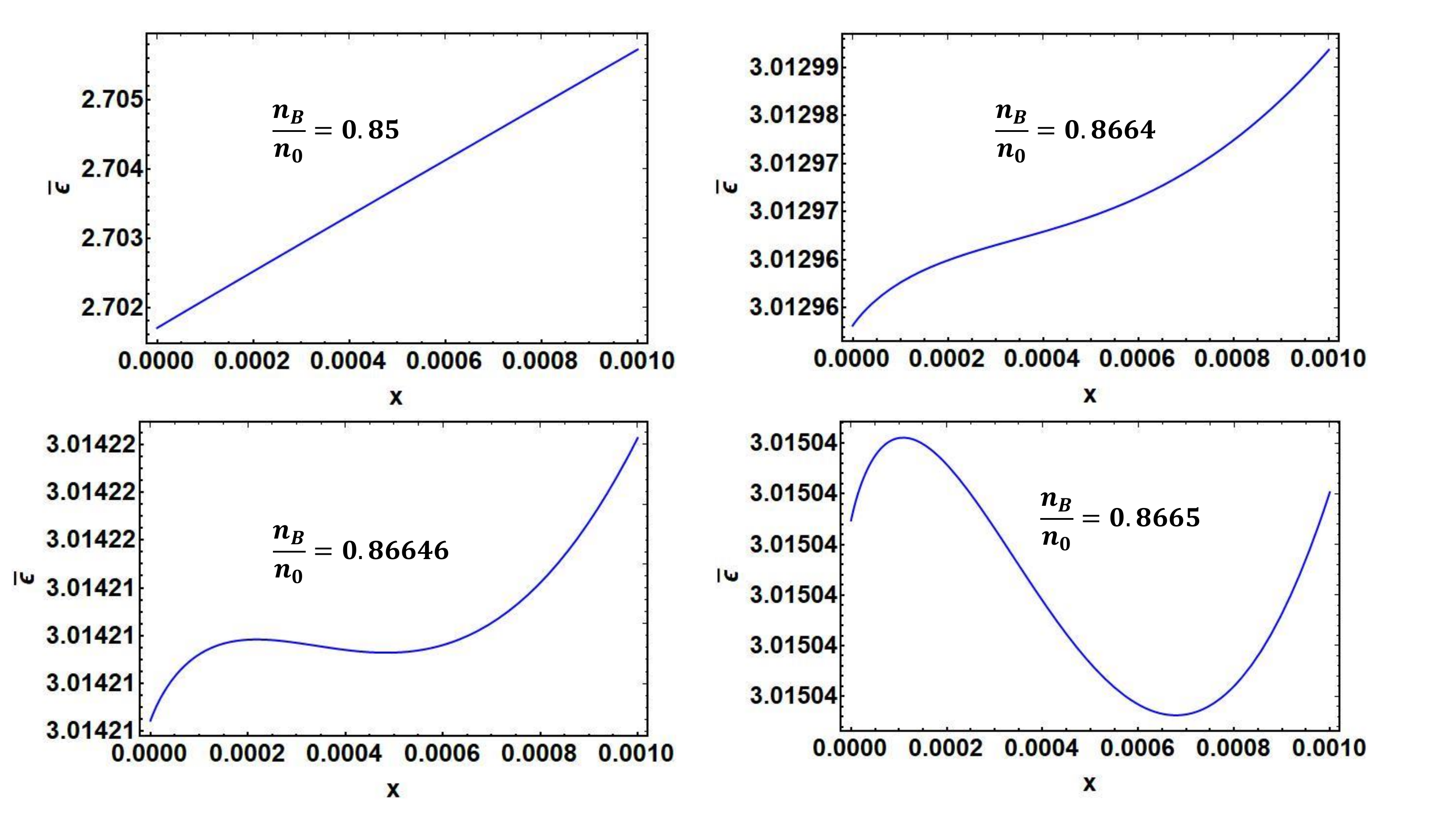}
	\caption{We plot the normalized energy density functional $\bar{\epsilon}$ of Eq. \ref{xye} as a function of the normalized quark density $x=n_B^Q/n_0$. The energy density functional of Eq. \ref{bar} exhibits a cuspy global minimum for $n_B^Q=0$ for small total baryon density. A local minimum appears at $n_B^Q\neq 0$ for higher baryon density which subsequently turns into a global minimum upon further increase in the baryon density.}
	\label{min2}
\end{figure}

Having minimized the energy density functional with respect to the quark density to obtain the equilibrium configuration, we can now proceed to compute the chemical potential which is given by 
\beq
\mu_B=\frac{d\epsilon_{\text{QSM}}}{dn_B},
\eeq
and the speed of sound using 
\beq
c_s^2=\frac{n_B}{\mu_B\frac{dn_B}{d\mu_B}},
\eeq
where $\epsilon_{\text{QSM}}$ is now evaluated at the quarkyonic configuration that minimizes its value. 
However, as noted in \cite{PhysRevC.101.035201} the simple model given by Eq. \ref{nex}, \ref{fermi1}, \ref{delta}, \ref{nq1}, \ref{e2} can violate causality and cause the speed of sound to turn negative. The reason behind this unphysical behavior is the abruptness with which the quarks appear when the quark onset takes place. To understand this we plot the energy density functional in Eq. \ref{e2} as a function of the quark density for a few values of the total baryon density. In order to do so efficiently we can define the following symbols 
\beq
x=\frac{n_B^Q}{n_0},\,\,\,\, y=\frac{n_B}{n_0},\,\,\,\,\bar{\epsilon}=5\pi^2 M_N\bigg(\frac{3\pi^2 n_0}{2}\bigg)^{-\frac{5}{3}}(\epsilon_{\text{QSM}} -M_N n_B) .
\label{xye}
\eeq
In terms of these variables the energy density in Eq. \ref{e2} can in the non-relativistic limit be expressed as
\beq
\bar{\epsilon}=\left(1-y+x\right)\left(\left(\frac{y-x}{1-y+x}+N_c^3x\right)^{5/3}-N_c^5x^{5/3}\right)+N_c^2x^{5/3}. 
\label{bar}
\eeq
 As is seen from the Figure \ref{min1}, for low values of the baryon density the minimum of the energy density lies at $x=n_B^Q=0$ and for higher density the minimum is at a higher quark density. Note that the minimum at $n_B^Q=0$ for low baryon density is a cuspy minimum. The abrupt transition from $x=n_B^Q=0$ to $n_B^Q\neq 0$, $x\neq 0$ is captured in Figure \ref{min2} and can be described as follows. Starting from a low value of baryon density at which there is only a single cuspy minimum at $x=n_B^Q=0$, as one increases the baryon density a local minimum appears at a finite but small  quark density. Upon increasing the density further this local minimum turns into a global minimum of the energy density functional. As a result the quark density goes from zero to a finite value abruptly. This is the onset density for baryons at which quarkyonic shell starts forming. This behavior may lead one to erroneously speculate that such a sudden change in the quark density is due to a first order phase transition from nuclear matter to quark matter. However, such speculation overlooks the key motivating factor behind using the quarkyonic model to describe matter at high density, which was precisely to avoid a phase transition between nuclear and quark matter. %In fact, as a confirmation, we can examine the energy density as a function of the baryon density for various values of the regulator $\Lambda$. As is clear from Figure \ref{e_QSM}, the energy density as a function of the baryon density is convex, ruling out the possibility of a first order phase transition. 

\begin{figure}
	\centering
	\includegraphics[width=\textwidth]{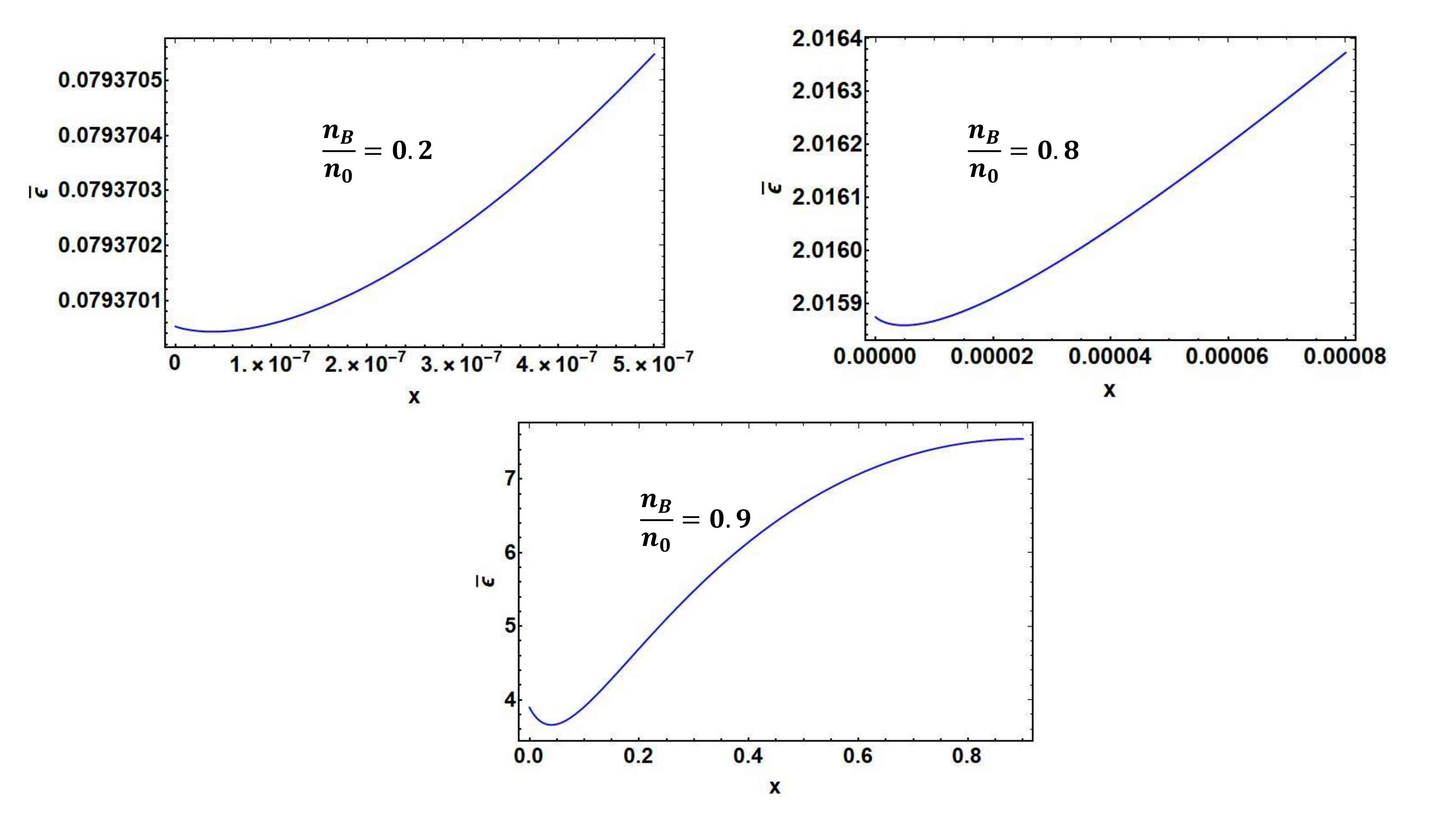}
	\caption{Normalized energy density functional with a finite $\Lambda$ plotted with increasing normalized quark density $x$ as defined in Eq. \ref{xye}. $\Lambda^3$ is chosen to be $10^{-5}\times3\pi^2n_0/2$.}
	\label{lmin}
\end{figure}

\begin{figure}
	\centering
		\includegraphics[width=0.5\textwidth]{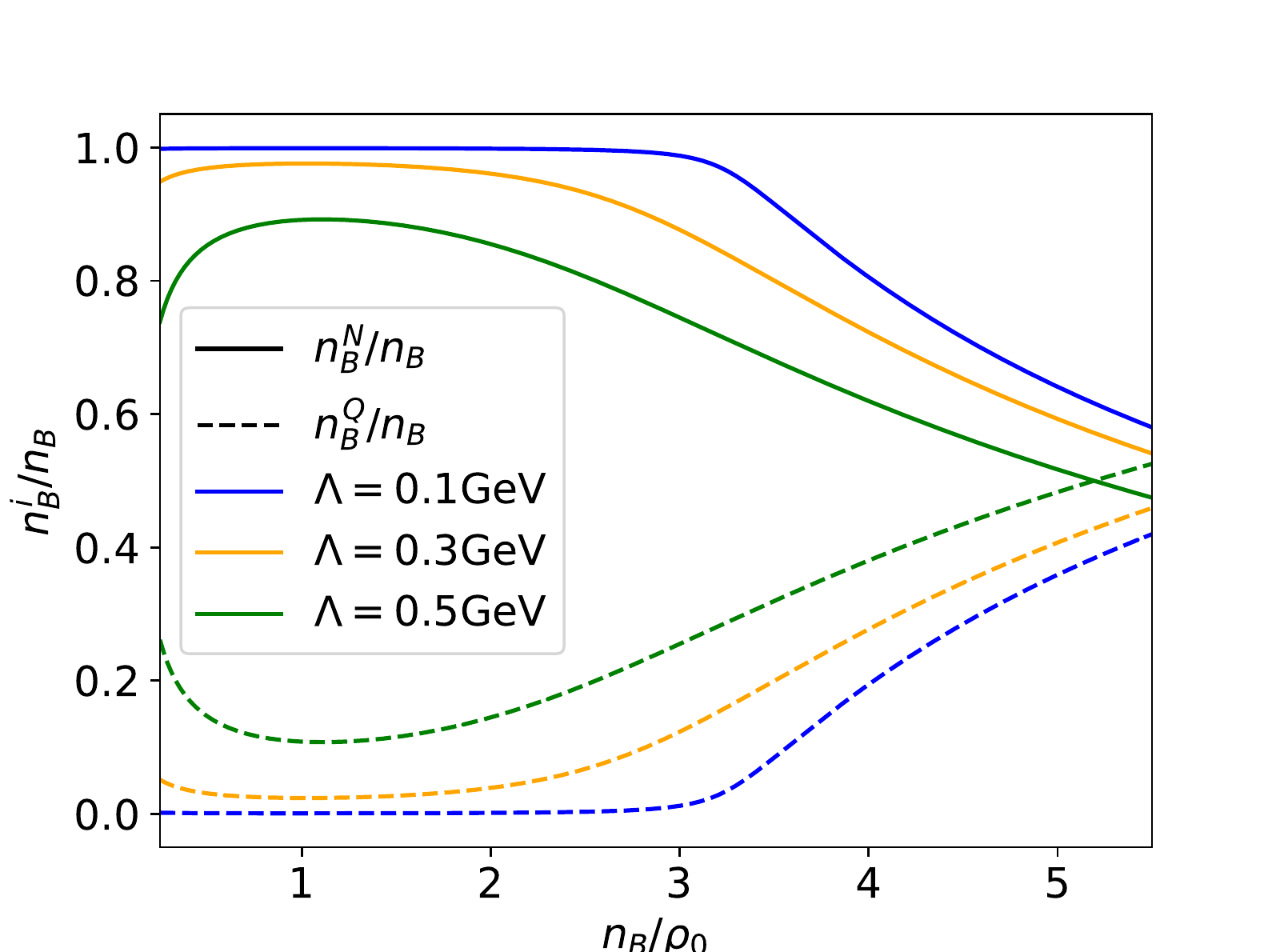}
		\caption{Nucleon density (solid lines) and quark density (dashed lines) for a few different values of regulator $\Lambda$ for the QSM model with hard core density $n_0 = 4\rho_0$ where $\rho_0$ is the nuclear saturation density.}	
		%	Neutron density (solid lines) and quark density (dashed lines) for a few different values of constant $\Lambda$. Hard core scattering density is set to $n_0 = 2\rho_0$}
		\label{regdense}
\end{figure}

The abruptness of the quark onset causes the density of baryons to change rather quickly for a modest change in the energy density. This in turn causes the chemical potential to drop with increasing density for a range of baryon densities which gives rise to unphysical behavior for the speed of sound. An attempt to remedy this was made in \cite{PhysRevC.101.035201} by modifying the quark density of states with 
\begin{figure}
	\centering
		\includegraphics[width=0.5\textwidth]{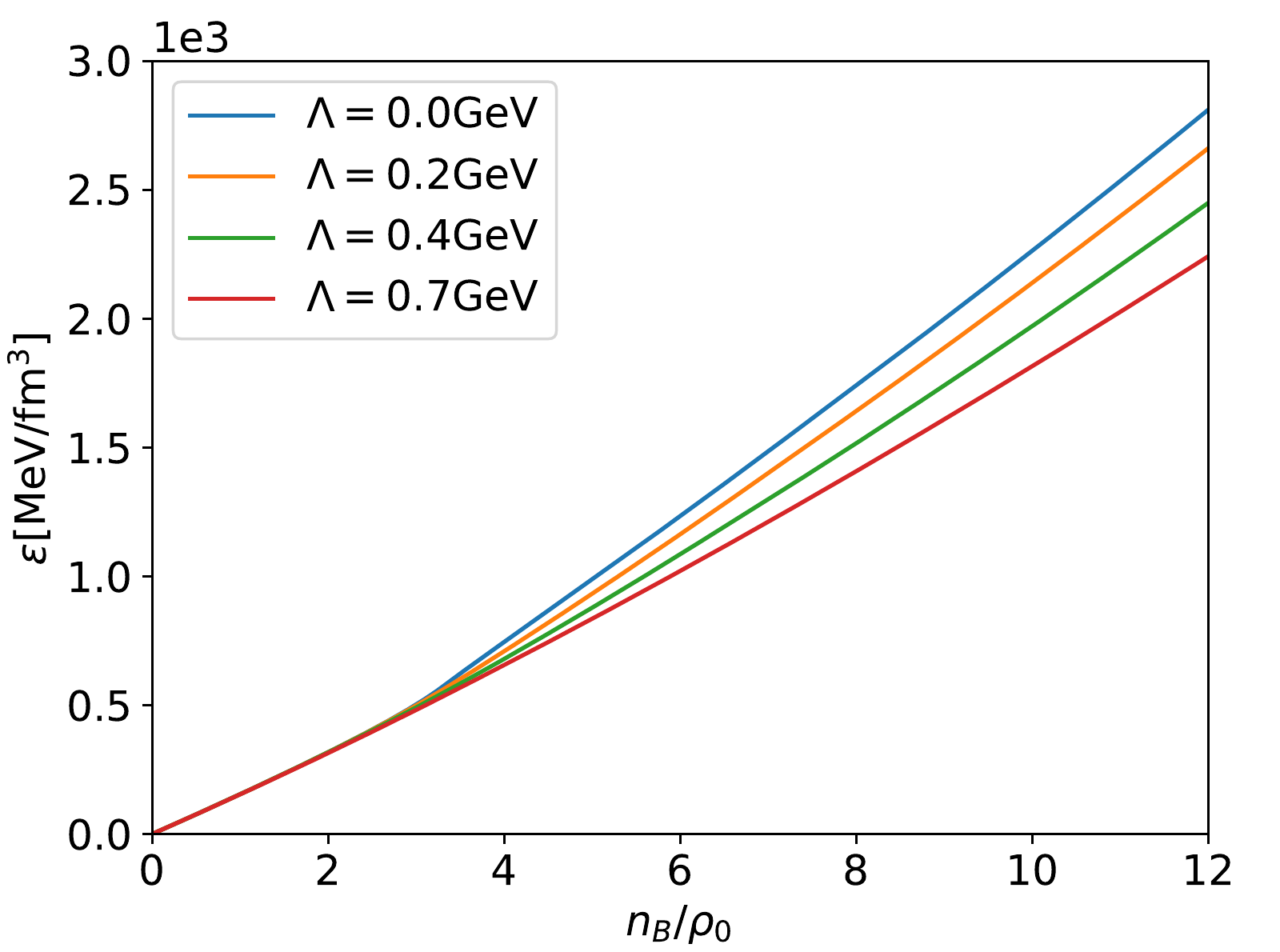}
	\caption{Minimized energy density $\epsilon$ as a function of baryon density $n_B$ for the QSM model. The hard core density is set at $n_0 = 4\rho_0$.}
	\label{e_QSM}
\end{figure}
\beq
g(k)=\frac{\sqrt{k^2+\Lambda^2}}{k},
\eeq
where $\Lambda$ is a regulator of mass dimension one.
In this case the quark density can be related to $k_F$ by
\beq
\frac{k_F}{N_c}=\sqrt{\left(\frac{3\pi^2}{2}n_B^Q+\Lambda^3\right)^{2/3}-\Lambda^2}.
\eeq
Also, the energy density functional is given by
\beq
\epsilon_{\text{QSM}}^{\Lambda}= 4\bigg(1-\frac{n_\text{B}^\text{N}}{n_0}\bigg)\int_{k_\text{F}}^{k_\text{F}+\Delta}\frac{\text{d}^3k}{(2\pi)^3}\sqrt{M_\text{N}^2+k^2}+4N_\text{C}\int_0^{k_\text{F}/N_\text{C}}\frac{d^3k}{(2\pi)^3}\sqrt{m^2+k^2}\left(\frac{\sqrt{k^2+\Lambda^2}}{|k|}\right).\nonumber\\
\label{e3}
\eeq
Defining an additional dimensionless variable $z$
\beq
z=\frac{\Lambda^3}{\frac{3\pi^2}{2}n_0},
\eeq
the energy density in Eq. \ref{e3} can be expressed in terms of $x, y, z$ as
\beq
\bar{\epsilon}&=&\left(1-y+x\right)\left(\left(\frac{y-x}{1-y+x}+N_c^3((x+z)^{2/3}-z^{2/3})^{3/2}\right)^{5/3}-N_c^5\left((x+z)^{2/3}-z^{2/3}\right)^{5/2}\right)\nonumber\\
&&\hspace{3in}+ N_c^2\left((x+z)^{5/3}-\frac{5}{3}z^{2/3}(x+z)+\frac{2}{3}z^{5/3}\right).\nonumber\\
\eeq
This modification results in a gradual appearance of quarks thus eliminating unphysical behavior in the chemical potential and the speed of sound as follows. The cuspy minimum of the energy density functional at $\Lambda=0$ for low baryon density turns into a cuspy maximum when $\Lambda\neq 0$ or $z\neq 0$. Also, the energy density functional for large enough $z$ has a single global minimum at a finite quark density for any nonzero baryon density. More specifically for baryon density much smaller than the hardcore density, the equilibrium quark density is small, but not exactly zero. As $n_B$ is increased further, the global minimum moves to higher quark densities reaching a significant fraction of the total baryon density once $n_B \sim n_0$ as seen in Figure \ref{lmin}. This eliminates the abrupt quark onset. In figure \ref{e_QSM} we plot the energy density to show that it is convex as a function of the density which confirms the absence of any first order phase transition, consistent with our expectation. However, the simple proposal for remedying the abruptness of quark onset by modifying the density of states \cite{PhysRevC.101.035201}, although effective in restoring causality, produces the undesired feature of having finite density of quarks at low values of the total baryon density. The amount of quarks present at low density is plotted in Figure \ref{regdense} for a few values of the parameter $\Lambda$. It is clear that higher values of $\Lambda$ results in higher quark densities at low baryon density. If we are to construct realistic models of quarkyonic matter we will have to devise a way to eliminate quarks at low density without reintroducing abruptness in quark onset. For example, one could consider a scenario where quark density is zero at low baryon density and then turns of gradually as the total baron density reaches some critical onset density. In the next section we will describe how such gradual onset of quarks can be realized at some critical onset density for baryons. With this we have now concluded the discussion of the excluded volume model for quarkyonic symmetric matter proposed in \cite{PhysRevC.101.035201}. 
\subsection{Modification to the QSM model}
\begin{figure}
	\centering
	\includegraphics[width=0.5\textwidth]{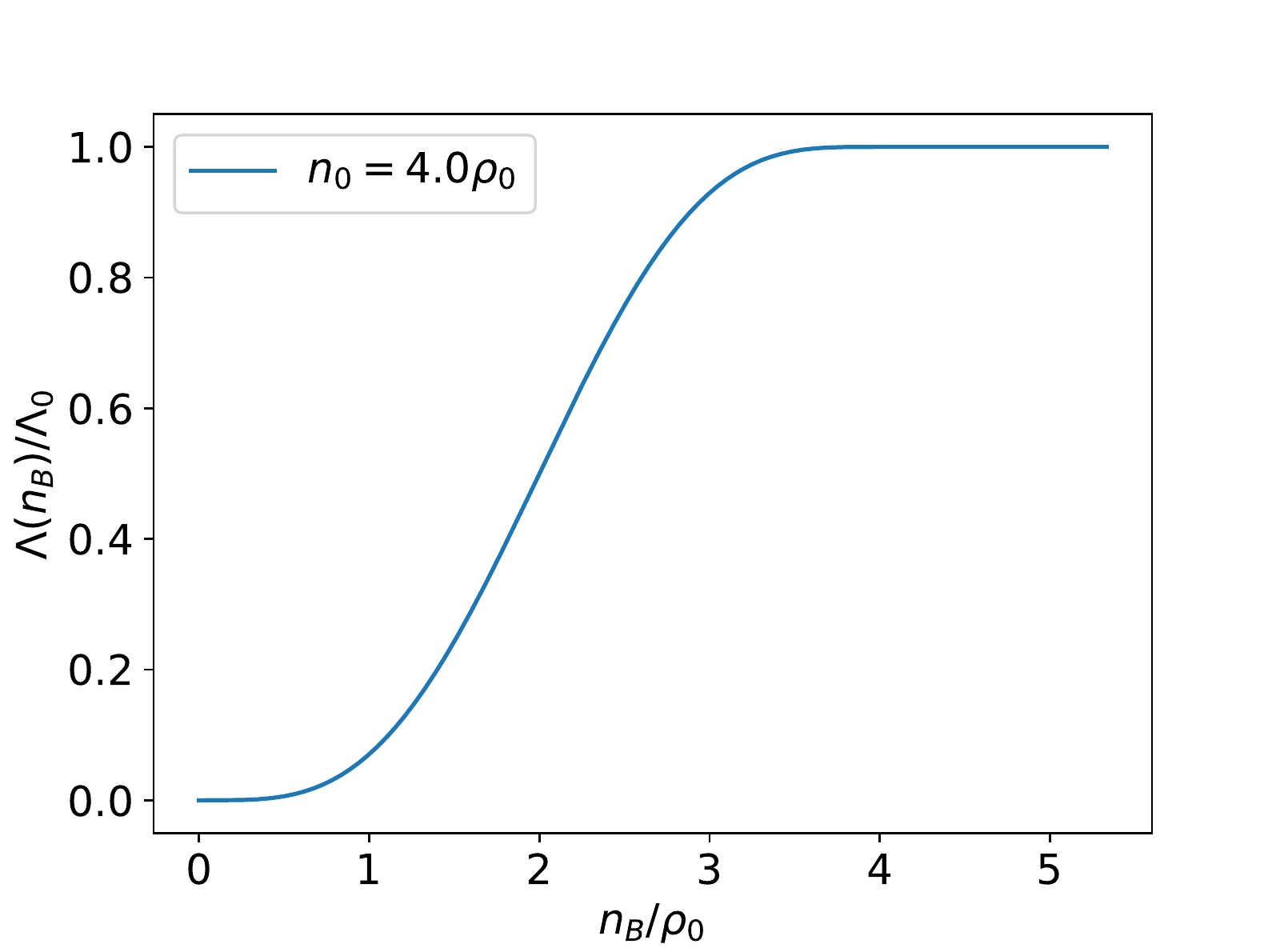}
	\caption{Density dependence of $\Lambda$ as discussed in the text using $n_0 = 4\rho_0$ where $\rho_0$ is the nuclear saturation density.}
	\label{Lambda(n)}
\end{figure}

\begin{figure}
	\centering
	\includegraphics[width=0.5\textwidth]{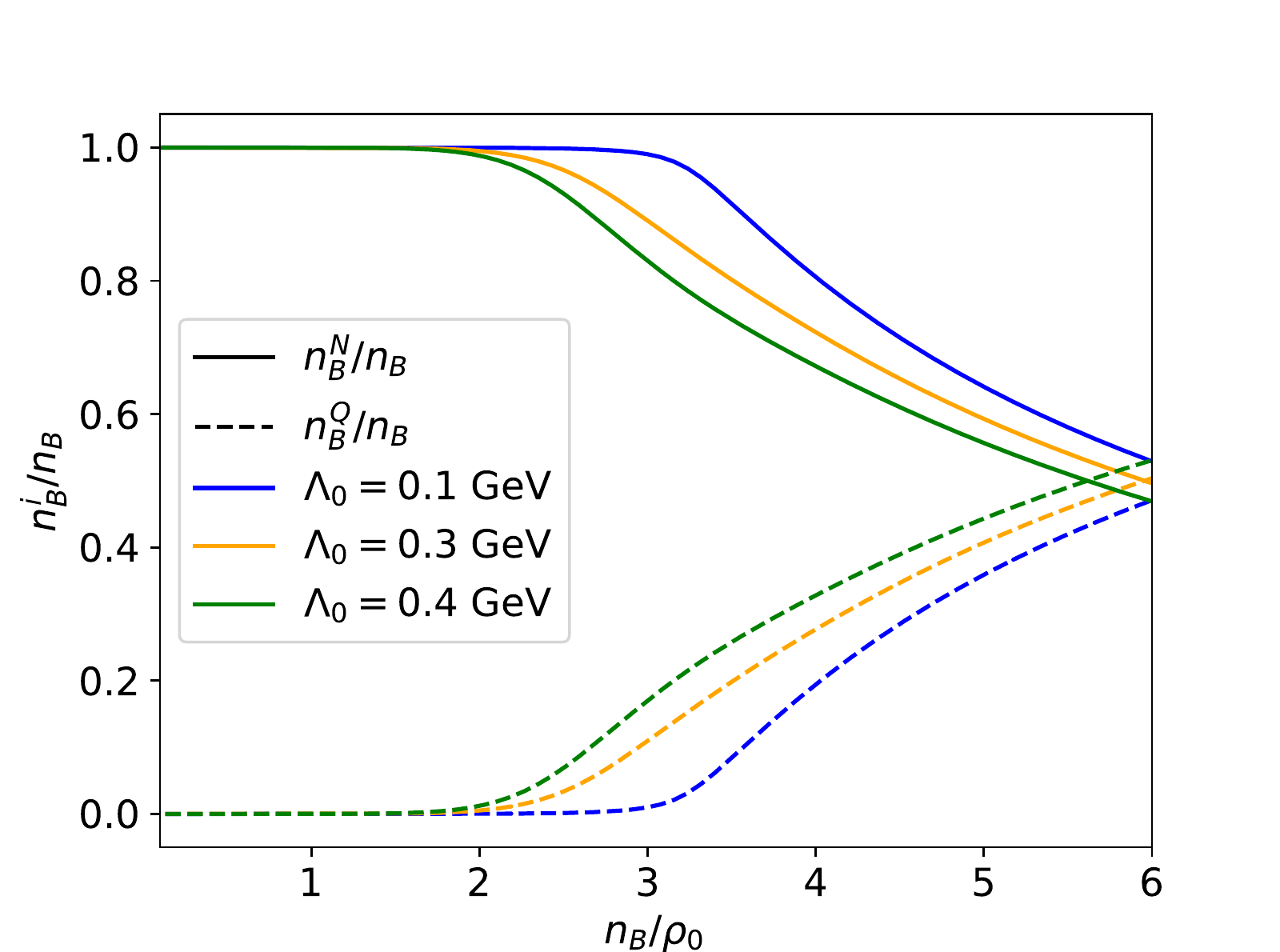}
	\caption{Nucleon density (solid lines) and quark density (dashed lines) for the QSM model with a density dependent $\Lambda(n_B)$. Hard core scattering density is set to $n_0 = 4\rho_0$ where $\rho_0$ is the nuclear saturation density.}
		%\textcolor{red}{Plot this for $n_0$ that you choose for the NS mass radius plots at the end}}
	\label{regdense2}
\end{figure}

As we observed in the previous subsection, the model proposed in \cite{PhysRevC.101.035201} with regulator for the quark density of states produces quarks for any $n_B\neq 0$. In what follows we will attempt to eliminate quarks completely for $n_B$ smaller than a critical baryon density. To achieve this behavior we alter the proposal in \cite{PhysRevC.101.035201} slightly by making the regulator $\Lambda$ density dependent. Note that, introducing density dependence for the regulator will only modify the quark contribution to the EOS. Since the regulator does not appear in the nuclear contribution to the energy density, it has no impact on the nuclear EOS except for eliminating quarks at low density. The density dependence of the regulator is motivated by the observation that when $\Lambda=0$ in the model of \cite{PhysRevC.101.035201} there exist no quarks at small total baryon density. Quarks begin to appear at low baryon density only when $\Lambda\neq 0$ and rather large. This implies that if we employ a density dependent regulator which has a small magnitude for low baryon density and is relatively large only at high baryon density, we will eliminate quarks at low baryon density while maintaining a gradual onset of quarks at higher baryon density. In Figure \ref{Lambda(n)} we illustrate a representative density dependence of $\Lambda$ which is able to achieve this. We choose the regulator or $\Lambda$ to be zero for $n_B=0$. %\hl{This is so as to make sure that the regulator does not interfere with nuclear physics.} 
$\Lambda$ is then increased with increasing density to reach some maximum value $\Lambda_0$ near the hardcore density beyond which it remains constant. 
The density dependence of the regulator in Figure \ref{Lambda(n)} can be expressed as a polynomial in $n_B^N/n_0$ as 
\beq
\Lambda(n_B^\text{N}) =\begin{cases}\Big[ -20\big(\frac{n_B^N}{n_0}\big)^7 +70\big(\frac{n_B^N}{n_0}\big)^6-84\big(\frac{n_B^N}{n_0}\big)^5+35\big(\frac{n_B^N}{n_0}\big)^4\Big]\Lambda_0& \text{for }n_B^N<n_0\\
\Lambda_0 &  \text{for }n_B^N\geq n_0 \end{cases}.
\label{lambda}
\eeq 
One can choose the functional form of the regulator to be different from what we chose so long as it smoothly and monotonically interpolates between $\Lambda(n_B^N=0)=0$ to $\Lambda(n_B^N=n_0)=\Lambda_0$ while keeping $d^n\Lambda/d(n_B^N)^n$ continuous for all values of $n_B$ for at least up to $n=3$. This is to ensure that the density dependence of the regulator does not introduce any discontinuities in the behavior of the chemical potential or the speed of sound. 
For the choice of the regulator in Figure \ref{Lambda(n)} the quark density is zero for small baryon densities and they gradually appear near some onset density as shown in Figure \ref{regdense2}. This onset density depends on both the hardcore density and the maximum value of the regulator $\Lambda_0$. For $\Lambda_0=0$, the onset density is very close to the hardcore density. For higher values of $\Lambda_0$ the onset density moves to lower values of the total baryon density as seen from Figure \ref{regdense2}. We will incorporate the density dependence of the regulator $\Lambda$ given in Eq. \ref{lambda} in the models of quarkyonic neutron matter that we construct next.

With this let us remind ourselves that the objective of this paper is to analyze mass and radius relations for the equations of state of quarkyonic matter within the excluded volume model. To have any hope of satisfying the neutron star mass radius constraints we will have to extend the quarkyonic model for isospin symmetric matter to a quarkyonic model whose hadronic sector is neutron rich. Such a model will also have to take into account electromagnetism and impose charge neutrality. The first step towards this is to construct a model which exhibits pure neutron matter at low density and gives rise to up and down quarks at high density where the up quark population is twice that of the down quark population to maintain charge neutrality. In the next section we construct these models. 

\section{quarkyonic description of neutron matter}  \label{3}
In this section we will construct excluded volume models for charge neutral quarkyonic matter with a hadronic sector that consists of neutrons only. We denote these models by QNM (quarkyonic neutron matter). Before we construct the QNM models, let us first review the constraints placed by neutron star observations and experiments on the equations of state in dense matter. In reference \cite{Essick:2020flb}, a combination of insights from gravitational wave observations of neutron star mergers and chiral perturbation theory helped constrain the radii of neutron stars with masses around 1.4 solar masses to $R_{1.4} = 11.4^{+1.38}_{-1.04}$km ($R^\text{NICER}_{1.4} = 12.54^{+0.71}_{-0.63}$km). Here and in what follows, constraints with superscript `NICER' includes data from the NICER measurement of PSR J0030+0451, while constraints without it does not. \cite{Essick:2020flb}\footnote{\cite{Essick:2020flb} improves on previously obtained constraints in \cite{Stringent_constraints}.} also imposed limits on the pressure at two and four times the saturation density, respectively $P_{2\text{sat}} = 14.2^{+18.1}_{-8.4}$MeV/fm$^3$ ($P_{2\text{sat}}^\text{NICER} = 28.7^{+15.3}_{-15.0}$MeV/fm$^3$) and $P_{4\text{sat}} = 202^{+87}_{-73}$MeV/fm$^3$ ($P_{4\text{sat}}^{\text{NICER}} = 211^{+108}_{-69}$MeV/fm$^3$). The pressure and symmetry energy at the saturation density, denoted as $P_0$ and $S_0$ respectively are further constrained by $1.9$MeV/fm$^3<P_0< 2.9$MeV/fm$^3$, $31.2\text{MeV}<S_0<34.3\text{MeV}$ \cite{PhysRevC.86.015803}. Similarly, we know from observations that the maximum mass of neutron stars is at least $\sim 2$ solar masses \cite{Cromartie2020,Antoniadis1233232}.

We will use the constraint on $P_0$ and $S_0$ to guide the construction of our excluded volume QNM models. We will then examine to what extent the equations of states of these models satisfy the remaining constraints mentioned in the previous paragraph. 

In constructing the QNM models, it is useful to recall how a non-dynamical model of charge neutral quarkyonic neutron matter proposed in \cite{McLerran:2018hbz} manages to satisfy the constraints on the pressure and symmetry energy at the saturation density. The model in \cite{McLerran:2018hbz} uses neutrons, up quarks and down quarks as quasiparticle degrees of freedom where the quarks are treated as free particles and nucleons interact among themselves via a nuclear potential of the form
\beq
V(n_B^N)=\tilde{a}n_B^N\frac{n_B^N}{\rho_0}+\tilde{b}n_B^N\left(\frac{n_B^N}{\rho_0}\right)^2.
\label{pot}
\eeq
As stated in \cite{McLerran:2018hbz} this potential is obtained from microscopic calculations of neutron matter in \cite{Gandolfi:2013baa, Gandolfi:2011xu}. 
The model in \cite{McLerran:2018hbz} uses $\tilde{a} \sim -28.8$ MeV, $\tilde{b}\sim 10.0$ MeV where $\rho_0$ is the nuclear saturation density $0.16$ fm$^{-3}$. Note that there is no hard-core excluded volume potential in this model. The corresponding energy density is given by
\beq
\epsilon_N&=&2\int_0^{k_{\text{Fermi}}} \frac{d^3k}{(2\pi)^3}\sqrt{k^2+M_N^2}+V(n_B^N) ,
\label{eLS}
\eeq
where 
\beq
n_B^N=\frac{k_{\text{Fermi}}^3}{3\pi^2},
\label{nLS}
\eeq
is the neutron density.
This produces a pressure of $2.4\,\text{MeV}/\text{fm}^3$ and a symmetry energy of $32 \text{MeV}$ at the saturation density which satisfy the constraints on $P_0$ and $S_0$. This non-dynamical model will guide the construction of excluded volume QNM models in the next section.

\subsection{The quarkyonic neutron matter models and incorporating nuclear interactions}
\label{nmqm}
We are now ready to construct excluded volume QNM (quarkyonic neutron matter) models where the hadronic sector includes only neutrons and the quark sector has two flavors: up and down quarks. The baryon density stored in neutrons is denoted as $n_B^N$ and that in up and down quarks is denoted as $n_B^u$ and $n_B^d$
such that
\beq
n_B^Q=n_B^u+n_B^d,\nonumber\\
n_B=n_B^N+n_B^Q.
\label{nqsplit}
\eeq
Note that, the symbol $n_B^N$ was used earlier to refer to nucleon density in the QSM model which had both neutrons and protons. In QNM models, we will use $n_B^N$ to stand for the density of neutrons since there are no protons in the model.
 The up and down quark fermi momenta are denoted as $k_u$ and $k_d$ respectively. Note that in the presence of a regulator $\Lambda$, the fermi momenta $k_u$ and $k_d$ are related to the baryon density stored in quarks by 
\beq
k_{u/d}=\sqrt{\left(3\pi^2n_B^{u/d}+\Lambda^3\right)^{2/3}-\Lambda^2}.
\label{q1}
\eeq
Additionally, charge neutrality forces 
\beq
n_B^d=2n_B^u,
\label{q2}
\eeq
which then relates $k_u$ and $k_d$. The fermi shell around quarks consists of neutron states only. The inner radius and the width of the neutron fermi shell are denoted by $k_F$ and $\Delta$ as before. Following \cite{McLerran:2018hbz} we set
\beq
k_d=\frac{k_F}{N_c}=\frac{k_F}{3}.
\label{q3}
\eeq
Neutron density $n_B^N$ is related to the Fermi momentum $k_F$ and the shell width $\Delta$ by
\beq
\frac{n_B^N}{1-\frac{n_B^N}{n_0}}=2\int_{k_F}^{k_F+\Delta}\frac{d^3k}{(2\pi)^3}
\label{fermi11}
\eeq
where we treat the neutrons as particles of a finite size with volume $1/n_0$. Comparing Eq. \ref{fermi1} to Eq. \ref{fermi11} , we can see a factor of $2$ difference in the degeneracy factors. This difference is to be attributed to the fact that in Eq. \ref{fermi1} we have two species of nucleons whereas in Eq. \ref{fermi11} we only consider neutrons. 
The energy density functional can then be written as 
\beq
\epsilon_{\text{QNM}} = 2\bigg(1-\frac{n_B^N}{n_0}\bigg)\int_{k_\text{F}}^{k_\text{F}+\Delta}\frac{\text{d}^3k}{(2\pi)^3}\sqrt{M_\text{N}^2+k^2}+\sum_{i=u,d}\frac{N_\text{C}}{\pi^2}\int_0^{k_\text{i}}k\,\text{d}k \sqrt{\Lambda^2+k^2}\sqrt{m^2+k^2},\nonumber\\
\label{q4}
\eeq
where we have taken equal up and down quark masses given by $m$. Using Eq. \ref{q1}, \ref{q2}, \ref{q3}, Eq. \ref{q4} can be expressed in terms of $n_B^u$ or $n_B^d$ alone. Minimizing the energy density functional with respect to $n_B^{u/d}$ produces the equilibrium configuration of quarkyonic matter for a particular total baryon density. The energy density for the equilibrium configuration at that baryon density can then be obtained by substituting the minimizing value of $n_B^{u/d}$ in Eq. \ref{q4}.\\ 
{\bf Introducing interactions:} In the energy density functional of Eq. \ref{q4} we so far have free neutrons except for the excluded volume effects which encode a hard core interaction. It is easy to see that such an energy density functional will not satisfy the constraints of pressure and symmetry energy at saturation density. %The corresponding mass and radius of neutron stars is plotted in \textcolor{red}{REf Fig}. In fact a nuclear interaction of the form
In order to see why this is the case let us compare the energy density of the equilibrium quarkyonic configuration obtained from Eq. \ref{q4} with the energy density of the model in \cite{McLerran:2018hbz} as given in Eq. \ref{eLS} at saturation density. 
%\beq
%V(n_B^N)=\tilde{a}n_B^N\frac{n_B^N}{\rho_0}+\tilde{b}n_B^N\left(\frac{n_B^N}{\rho_0}\right)^2.
%\label{pot}
%\eeq
%where $\rho_0$ is the nuclear saturation density, was used in [\textcolor{red}{CITE Sanjay Larry}] to reproduce the constraints of pressure and symmetry energy for the quarkyonic model there.  To match these constraints of pressure and energy density we will have to modify the excluded volume model further. To motivate these modifications we first write down the dependence of the energy density of interacting neutron matter on neutron density in \textcolor{red}{[CITE LS]} in the absence of any quarks 
The energy density in \cite{McLerran:2018hbz} for low baryon density in the absence of quarks, following Eq. \ref{eLS} is given by 
\beq
\epsilon_N&=&2\int \frac{d^3k}{(2\pi)^3}\sqrt{k^2+M_N^2}+V(n_B^N)\nonumber\\
&\sim & 2\int \frac{d^3k}{(2\pi)^3}\left(M_N+\frac{k^2}{2M_N}\right)+V(n_B^N)+\cdots\nonumber\\
&\sim & M_Nn_B^N +\frac{(3\pi^2n_B^N)^{5/3}}{10\pi^2 M_N}+\tilde{a}n_B^N\frac{n_B^N}{\rho_0}+\tilde{b}n_B^N\left(\frac{n_B^N}{\rho_0}\right)^2+\cdots,
\label{LS}
\eeq
where the neutron Fermi momentum is related to the neutron density according to Eq. \ref{nLS}. 
We also take the non-relativistic expansion in the second line of Eq. \ref{LS} and the ellipsis stand for higher order terms in the non-relativistic expansion. In absence of quarks the energy density of neutrons in the excluded volume is on the other hand
\beq
\epsilon_{\text{QNM}}&=&2\bigg(1-\frac{n_B^N}{n_0}\bigg)\int_{0}^{\Delta}\frac{\text{d}^3k}{(2\pi)^3}\sqrt{M_\text{N}^2+k^2}\nonumber\\
&\sim & M_N n_B^N+\frac{\left(3\pi^2n_B^N\right)^{5/3}}{10\pi^2 M_N}\left(1-\frac{n_B^N}{n_0}\right)^{-2/3}+\cdots\nonumber\\
&\sim & M_N n_B^N+\frac{\left(3\pi^2n_B^N\right)^{5/3}}{10\pi^2 M_N}+\frac{\left(3\pi^2n_B^N\right)^{5/3}}{10\pi^2 M_N}\frac{2n_B^N}{3n_0}+\cdots,
\label{SS}
\eeq
where in the second line of Eq. \ref{SS} we have employed the non-relativistic limit and in the final step we have expanded in the parameter $n_B^N/n_0$ for small $n_B^N/n_0$. Since the hardcore density sets the baryon density at which quarks appear, the hardcore density should be larger than the nuclear saturation density i.e. $n_0>\rho_0$ making the expansion in small $n_B^N/n_0$ useful near $n_B^N \sim \rho_0$. Comparing Eq. \ref{LS} and \ref{SS} we see that the behaviors of the energy density as a function of the neutron density  are very different in the two cases for $n_B^N\sim\rho_0$. Hence, there is no reason to expect the excluded volume model of Eq. \ref{q4} to satisfy the constraints of pressure and the symmetry energy in the vicinity of $n_B^N\sim \rho_0$. 

In order to remedy this, we have to modify the excluded volume model further. Before we do this let us first devise a working definition of the nuclear potential in the excluded volume model labeled `i' the following way
\beq
V_{\text{i}}=\epsilon_\text{i}(n_B^Q=0) - 2\int\frac{d^3k}{(2\pi)^3}\sqrt{k^2+M_N^2}=\epsilon_\text{i}(n_B^Q=0) -\int_0^{k_{\text{Fermi}}}\frac{k^2\sqrt{k^2+M_N^2}}{\pi^2},
\label{potdef}
\eeq
where $k_{\text{Fermi}}=(3\pi^2 n_B^N)^{1/3}$, $\epsilon_\text{i}(n_B^Q=0)$ is the energy density functional of the excluded volume model `i' when the quark density is set to zero. According to this definition, the nuclear potential of the excluded volume model of QNM in Eq. \ref{q4} is given by
\beq
V_{\text{QNM}}(n_B^N)&=&\frac{2}{3}\frac{\left(3\pi^2n_B^N\right)^{5/3}}{10\pi^2 M_N}\frac{n_B^N}{n_0}+\frac{5}{9}\frac{\left(3\pi^2n_B^N\right)^{5/3}}{10\pi^2 M_N}\left(\frac{n_B^N}{n_0}\right)^2\nonumber\\
&&+\frac{40}{81}\frac{\left(3\pi^2n_B^N\right)^{5/3}}{10\pi^2 M_N}\left(\frac{n_B^N}{n_0}\right)^3+\frac{110}{243}\frac{\left(3\pi^2n_B^N\right)^{5/3}}{10\pi^2 M_N}\left(\frac{n_B^N}{n_0}\right)^4\cdots,\nonumber\\
\label{q4a}
\eeq
where we have taken the nonrelativistic limit and expanded in $n_B^N/n_0$. The potential in Eq. \ref{q4a} is very different from the potential in Eq. \ref{pot} near the saturation density which explains why we don't expect the energy density functional of the QNM model to satisfy the constraints of pressure and symmetry energy at the saturation density. We want the nuclear interactions in the excluded volume model to mimic the nuclear potential in Eq. \ref{pot}. The first step towards achieving this is to add the nuclear potential of Eq. \ref{pot} to the energy density functional of QNM model of Eq. \ref{q4} and define
\beq
\epsilon_{\text{QNMV}} = \epsilon_{\text{QNM}}+V(n_B^N).\nonumber\\
\label{q42}
\eeq 
%\beq
%\epsilon_{\text{NMQMV}} = 2\bigg(1-\frac{n_B^N}{n_0}\bigg)\int_{k_\text{F}}^{k_\text{F}+\Delta}\frac{\text{d}^3k}{(2\pi)^3}\sqrt{M_\text{N}^2+k^2}+V(n_B^N)+\sum_{j=u,d}\frac{N_\text{C}}{\pi^2}\int_0^{k_\text{j}}k\,\text{d}k \sqrt{\Lambda^2+k^2}\sqrt{m^2+k^2}.\nonumber\\
%\label{q42}
%\eeq 
Again, in the absence of quarks and in the non-relativistic limit for the nucleons, the nuclear potential in Eq. \ref{q42} is of the form
\beq
V_{\text{QNMV}}(n_B^N)&=&\tilde{a}n_B^N\frac{n_B^N}{\rho_0}+\tilde{b}n_B^N\left(\frac{n_B^N}{\rho_0}\right)^2+\frac{2}{3}\frac{\left(3\pi^2n_B^N\right)^{5/3}}{10\pi^2 M_N}\frac{n_B^N}{n_0}+\frac{5}{9}\frac{\left(3\pi^2n_B^N\right)^{5/3}}{10\pi^2 M_N}\left(\frac{n_B^N}{n_0}\right)^2\nonumber\\
&&+\frac{40}{81}\frac{\left(3\pi^2n_B^N\right)^{5/3}}{10\pi^2 M_N}\left(\frac{n_B^N}{n_0}\right)^3+\frac{110}{243}\frac{\left(3\pi^2n_B^N\right)^{5/3}}{10\pi^2 M_N}\left(\frac{n_B^N}{n_0}\right)^4\cdots\nonumber\\
&=&\tilde{a}n_B^N\left(\frac{n_B^N}{\rho_0}\right)+\tilde{b}n_B^N\left(\frac{n_B^N}{\rho_0}\right)^2+\left(\frac{2}{3}\frac{\left(3\pi^2n_B^N\right)^{5/3}}{10\pi^2 M_N}\frac{\rho_0}{n_0}\right)\left(\frac{n_B^N}{\rho_0}\right)\nonumber\\
&+&\left(\frac{5}{9}\frac{\left(3\pi^2n_B^N\right)^{5/3}}{10\pi^2 M_N}\left(\frac{\rho_0}{n_0}\right)^2\right)\left(\frac{n_B^N}{\rho_0}\right)^2+\left(\frac{40}{81}\frac{\left(3\pi^2n_B^N\right)^{5/3}}{10\pi^2 M_N}\left(\frac{\rho_0}{n_0}\right)^3\right)\left(\frac{n_B^N}{\rho_0}\right)^3\nonumber\\
&+&\left(\frac{110}{243}\frac{\left(3\pi^2n_B^N\right)^{5/3}}{10\pi^2 M_N}\left(\frac{\rho_0}{n_0}\right)^4\right)\left(\frac{n_B^N}{\rho_0}\right)^4+\cdots
\label{q43}
\eeq
In Eq. \ref{q43} we have also expanded in $n_B^N/n_0$.
However the potential in Eq. \ref{q43} deviates significantly from the potential in Eq. \ref{pot} 
near the saturation density $n_B^N\sim\rho_0=0.16 \text{fm}^{-3}$. For $n_B^N=\rho_0$, the potential in Eq. \ref{pot} or the first two terms on the RHS of Eq. \ref{q43} evaluate to 
\beq
V(\rho_0)=\tilde{a}\rho_0+\tilde{b}\rho_0\sim -28.8 \rho_0 + 10.0\rho_0,
\eeq 
where we have substituted $\tilde{a}=-28.8$ MeV and $\tilde{b}=10.0$ MeV. The third term in the RHS of Eq. \ref{q43} evaluates to
\beq
\left(\frac{2}{3}\frac{\left(3\pi^2n_B^N\right)^{5/3}}{10\pi^2 M_N}\frac{\rho_0}{n_0}\right)\left(\frac{n_B^N}{\rho_0}\right)\sim 23.5 \rho_0 \left(\frac{\rho_0}{n_0}\right).
\eeq
As a result, Eq. \ref{q43} will not match the constraints in its current form unless $\rho_0\ll n_0$. It is now clear that the contribution to the energy density functional coming from the excluded volume counterpart has to be much smaller than it is in Eq. \ref{q43} near the saturation density. There are two different approaches to engineer this as we demonstrate below.
\subsubsection{Approach I}
The first approach goes as follows. To obtain the QNMV model in Eq. \ref{q42} we had added the potential in Eq. \ref{pot} to the energy density of the QNM model in Eq. \ref{q4}. We will now modify the potential in Eq. \ref{pot} in such a way that when added to the energy density of Eq. \ref{q4}, the corresponding nuclear potential as defined in Eq. \ref{potdef} will match with the potential in Eq. \ref{pot} for densities less than or equal to the saturation density $\rho_0$ up to corrections that keep the pressure and symmetry energy within bounds set by constraints at $\rho_0$. In order to do this we first truncate the expansion in Eq. \ref{q43} up to a certain order $n$ in $n_B^N/n_0$ and denote the corresponding potential as $V_\text{QNMV}^{n}$. We can then subtract off $V_\text{QNMV}^{n}-V(n_B^N)$ from $V_{\text{QNMV}}$, where $V(n_B^N)$ is the potential in Eq. \ref{pot} to get the modified expression for the potential that replaces Eq. \ref{pot}. For example, if we truncate after second order in the expansion $n_B^N/n_0$ we get the potential
\beq
\tilde{V}(n^N_B)&=&V(n_B^N)-\left(\frac{2}{3}\frac{\left(3\pi^2n_B^N\right)^{5/3}}{10\pi^2 M_N}\frac{\rho_0}{n_0}\right)\left(\frac{n_B^N}{\rho_0}\right)-\left(\frac{5}{9}\frac{\left(3\pi^2n_B^N\right)^{5/3}}{10\pi^2 M_N}\left(\frac{\rho_0}{n_0}\right)^2\right)\left(\frac{n_B^N}{\rho_0}\right)^2.\nonumber\\
%&&-\left(\frac{40}{81}\frac{\left(3\pi^2n_B^N\right)^{5/3}}{10\pi^2 M_N}\left(\frac{\rho_0}{n_0}\right)^3\right)\left(\frac{n_B^N}{\rho_0}\right)^3-\left(\frac{110}{243}\frac{\left(3\pi^2n_B^N\right)^{5/3}}{10\pi^2 M_N}\left(\frac{\rho_0}{n_0}\right)^4\right)\left(\frac{n_B^N}{\rho_0}\right)^4.\nonumber\\
\label{pot3}
\eeq
We can now write down the energy density functional for the corresponding excluded volume model by adding the potential in Eq. \ref{pot3} to the energy density in Eq. \ref{q4} as 
\beq
\epsilon_{\text{QNMV}1}=\epsilon_{\text{QNM}}+\tilde{V}(n_N^B).
\label{NMV1}
\eeq
We will refer to this model as QNMV1.
The idea behind truncating the expansion in Eq. \ref{q43} and defining $\tilde{V}$ is as follows. We want the effect of the excluded volume potential to significantly affect the energy density only for the densities that are high compared to the saturation density. Near the saturation density, we want the the contribution of the nuclear potential in Eq. \ref{pot} to dominate over that from the excluded volume counterpart so that the equation of state satisfies the constraints on the pressure and symmetry energy at low density. Hence, we truncate the expansion of Eq. \ref{q43} only up to an order which keeps the energy and pressure at saturation density within the experimental constraints. We find that this can be achieved if we truncate the expansion at order $n=2$.

\subsubsection{Approach II}
\begin{figure}
	\centering
	\includegraphics[width=0.5\textwidth]{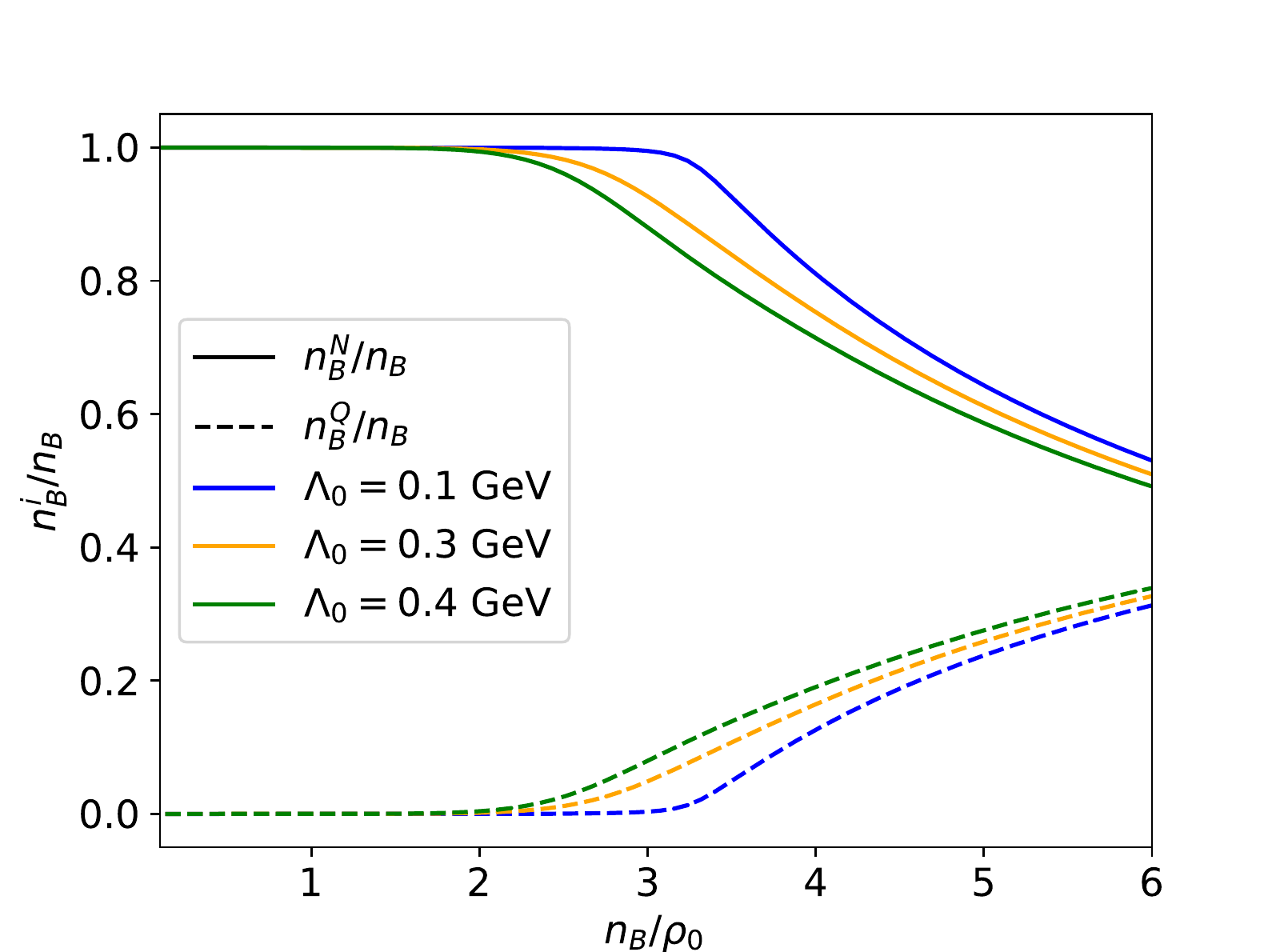}
	\caption{Nucleon density (solid lines) and quark density (dashed lines) for the QNM model with a density dependent $\Lambda(n_B)$. Hard core scattering density is set to $n_0 = 4\rho_0$ where $\rho_0$ is the nuclear saturation density.}
	%\textcolor{red}{Plot this for $n_0$ that you choose for the NS mass radius plots at the end}}
	\label{popdensQNM}
\end{figure}

The second approach to writing down an excluded volume model that satisfies the constraints of pressure and symmetry energy at saturation density is to generalize the excluded density defined in Eq. \ref{nex}. This generalization has to be constructed in such a way that it minimizes the contribution of the excluded volume hard core potential near the saturation density while also imposing an upper bound on the neutron density. For this to work, of course, the upper bound on neutron density or the hard core density has to be larger than the nuclear saturation density. The ansatz for the excluded density stated in Eq. \ref{nex} is only one of the possible choices for a potential that imposes an upper bound on nucleon/neutron density. A more general form of the potential can be found by postulating the following relation between the excluded density of neutrons and the actual density of neutrons
\beq
n_{\text{ex}}^N=\frac{n_B^N}{1-\left(\frac{n_B^N}{n_0}\right)^{\gamma}},
\eeq
where $\gamma$ is now a parameter. The original excluded volume model for quarkyonic matter as defined in Eq. \ref{nex} sets $\gamma=1$. In our modification to the excluded volume we will choose $\gamma>1$ so as to suppress the effects of the interaction terms contributed by the excluded volume potential near the saturation density. To see how this works, let's first write down the energy density of our modified excluded volume model defined as 
\beq
\epsilon_{\text{QNMV}2} &=& 2\Bigg[1-\bigg(\frac{n_B^N}{n_0}\bigg)^\gamma\Bigg]\Bigg[\int_{k_\text{F}}^{k_\text{F}+\Delta}\frac{\text{d}^3k}{(2\pi)^3}\sqrt{M_\text{N}^2+k^2}\Bigg]+V(n^N_B)\nonumber\\
&+&\sum_{i=u,d}\frac{N_\text{C}}{\pi^2}\int_0^{k_i}\text{d}k \sqrt{\Lambda^2+k^2}\sqrt{m_\text{Q}^2+k^2}.\nonumber\\
\label{e1}
\eeq 
We can now extract the nuclear potential of this model using Eq. \ref{potdef}  
\beq
V_{\text{QNMV}2}=\tilde{a}n_B^N\frac{n_B^N}{\rho_0}+\tilde{b}n_B^N\left(\frac{n_B^N}{\rho_0}\right)^2+\frac{\left(3\pi^2n_B^N\right)^{5/3}}{10\pi^2 M_N}\frac{2}{3}\left(\frac{n_B^N}{n_0}\right)^{\gamma}+\frac{\left(3\pi^2n_B^N\right)^{5/3}}{10\pi^2 M_N}\frac{5}{9}\left(\frac{n_B^N}{n_0}\right)^{2\gamma}\cdots\nonumber\\
\eeq
where we have carried out a non-relativistic expansion in the neutron mass and also expanded in $n_B^N/n_0$. It is now clear that we can suppress the effect of the excluded volume contributions for $n_B^N\sim \rho_0<n_0$ by choosing $\gamma>1$. We denote this model as QNMV2. In figure \ref{popdensQNM} we show the population density for QNM plotted for $n_0 = 4\rho_0$. We note that the quark onset densities are similar to those of QSM, but the portion of quarks is somewhat smaller. The population density for QNMV1 and QNMV2 are almost identical to that of QNM, and so we did not find it necessary to include a plot of those in this paper.  

Note that in the models we constructed the nuclear potential persists in the energy density functional at high density when quarks appear. Since we do not know much about nuclear interactions at such high density, it may be judicious to turn off the nuclear interactions at these densities. However, we refrain from making this choice since at densities above $\sim 0.75n_0$ it is found that the difference in the energy density and pressure with and without the interactions is less than 10\%.

Having obtained the various models that we discussed so far we can now examine which of these models violate the experimental and observational constraints obtained from neutron star observations.
\begin{figure}[h!]
\centering
\includegraphics[width=.5\textwidth]{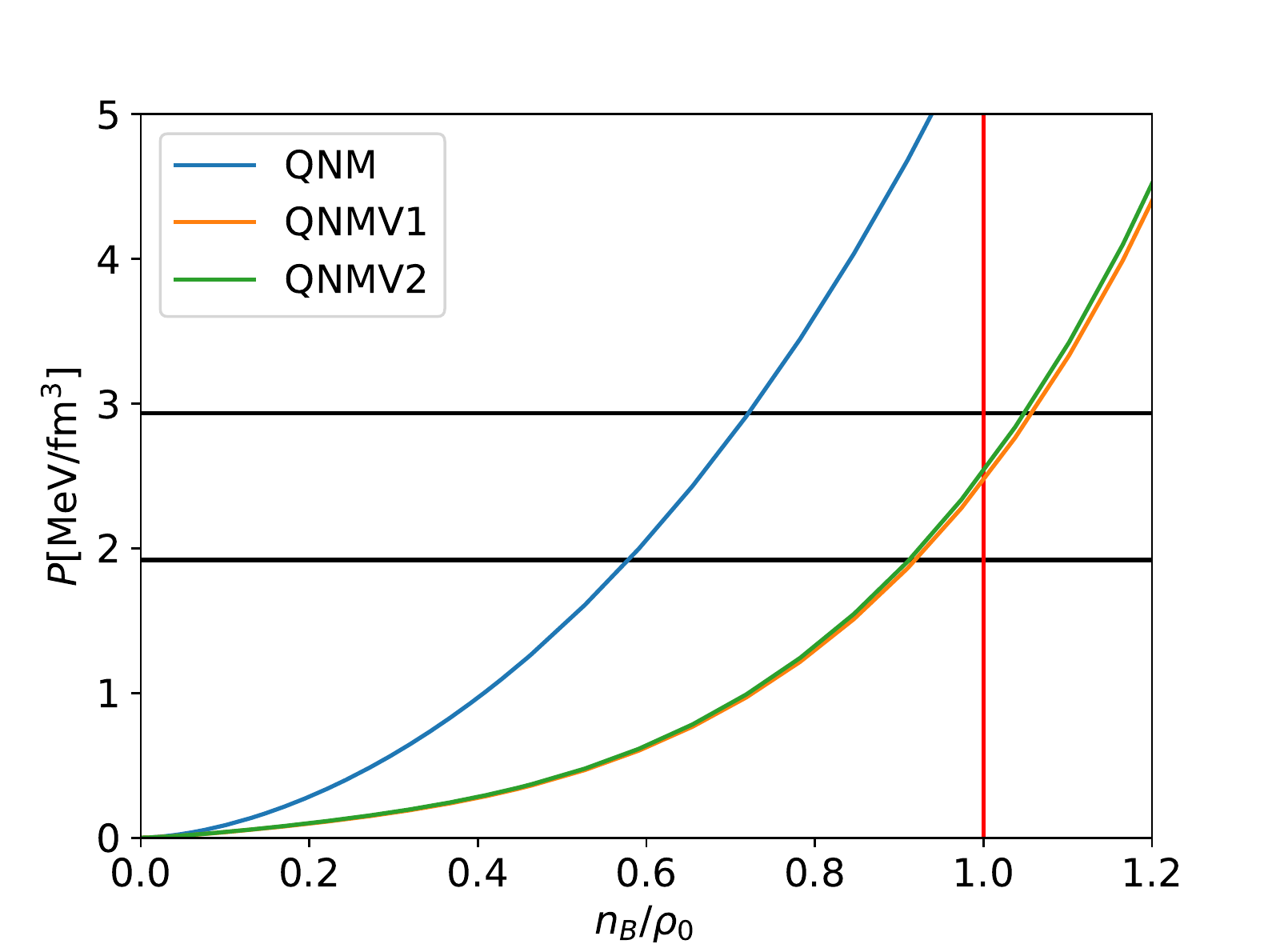}
\caption{We plot the pressure around the saturation density for the three models QNM, QNMV1 and QNMV2. The saturation density $\rho_0$ is indicated by a vertical red line, while the upper and lower bounds on the pressure at the saturation density are shown by horizontal black lines.  For QNM we use the parameter values $n_0 = 4.5\rho_0$ and $\Lambda_0 = 0.25$GeV. For QNMV1, we use $n_0=4\rho_0$ and $\Lambda_0=0.25$GeV. For QNMV2 we use $n_0=4\rho_0$, $\Lambda_0=0.3$GeV and $\gamma=2$. We choose $\tilde{a}=-28.6$ MeV and $\tilde{b}=10$ MeV for QNMV1 and $\tilde{a}=-27.6$ MeV and $\tilde{b}=7.9$ MeV for QNMV2.}
	%Some examples of the pressure around saturation density realized by the original model \cite{PhysRevC.101.035201} (OM), OM with included quark degrees of freedom (OM+QDF), OM+QDF with nuclear potential (OM+QDF+pot) and OM+QDF+pot with the $\gamma$ factor included as discussed in the text (OM+QDF+pot+$\gamma$). Horizontal lines mark the upper and lower experimental bounds at saturation density as proposed in \cite{PhysRevLett.108.081102}.}
\label{Pressure around saturation density}
\end{figure}
\begin{figure}
	\centering
	\includegraphics[width=0.5\textwidth]{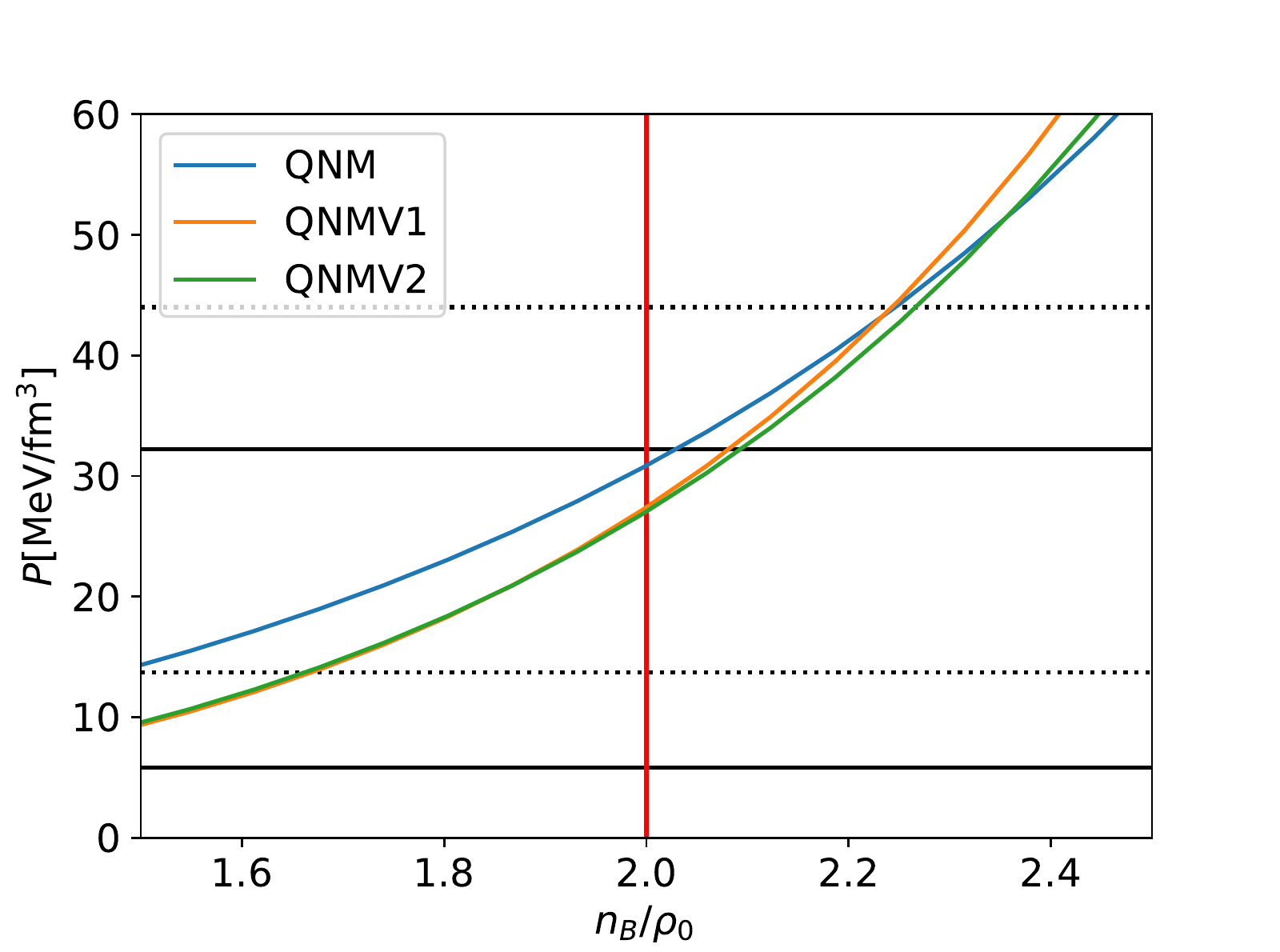}
	\caption{We plot the pressure around two times the saturation density a function for the three models QNM, QNMV1 and QNMV2. For QNM we use the parameter values $n_0 = 4.5\rho_0$ where $\rho_0$ is the nuclear saturation density and $\Lambda_0 = 0.25$GeV. For QNMV1, we use $n_0=4\rho_0$ and $\Lambda_0=0.25$GeV. For QNMV2 we use, $n_0=4\rho_0$, $\Lambda_0=0.3$GeV and $\gamma=2$. Also, we choose $\tilde{a}=-28.6$ MeV and $\tilde{b}=10$ MeV for QNMV1 and $\tilde{a}=-27.6$ MeV and $\tilde{b}=7.9$ MeV for QNMV2. The red vertical line represents the density $n_B = 2\rho_0$, while the upper and lower bounds on the pressure at $n_B = 2\rho_0$ excluding (including) the NICER measurement of PSR J0030+0451 is shown by horizontal black solid (dashed) lines.}%, and the solid black line indicates the maximum allowed pressure inside neutron stars.}%Some examples of the pressure as a function of baryon density. Legends are as in figure \ref{Pressure around saturation density}. Dashed horizontal lines shows the upper and lower exprimental bounds on the pressure at 4 times the saturation density, and solid horizontal line shows the experimental constraint on the maximum pressure in neutron stars, as proposed in \cite{Stringent_constraints}. }
	\label{pressure around two times saturation}
\end{figure}
\begin{figure}
	\centering
	\includegraphics[width=0.5\textwidth]{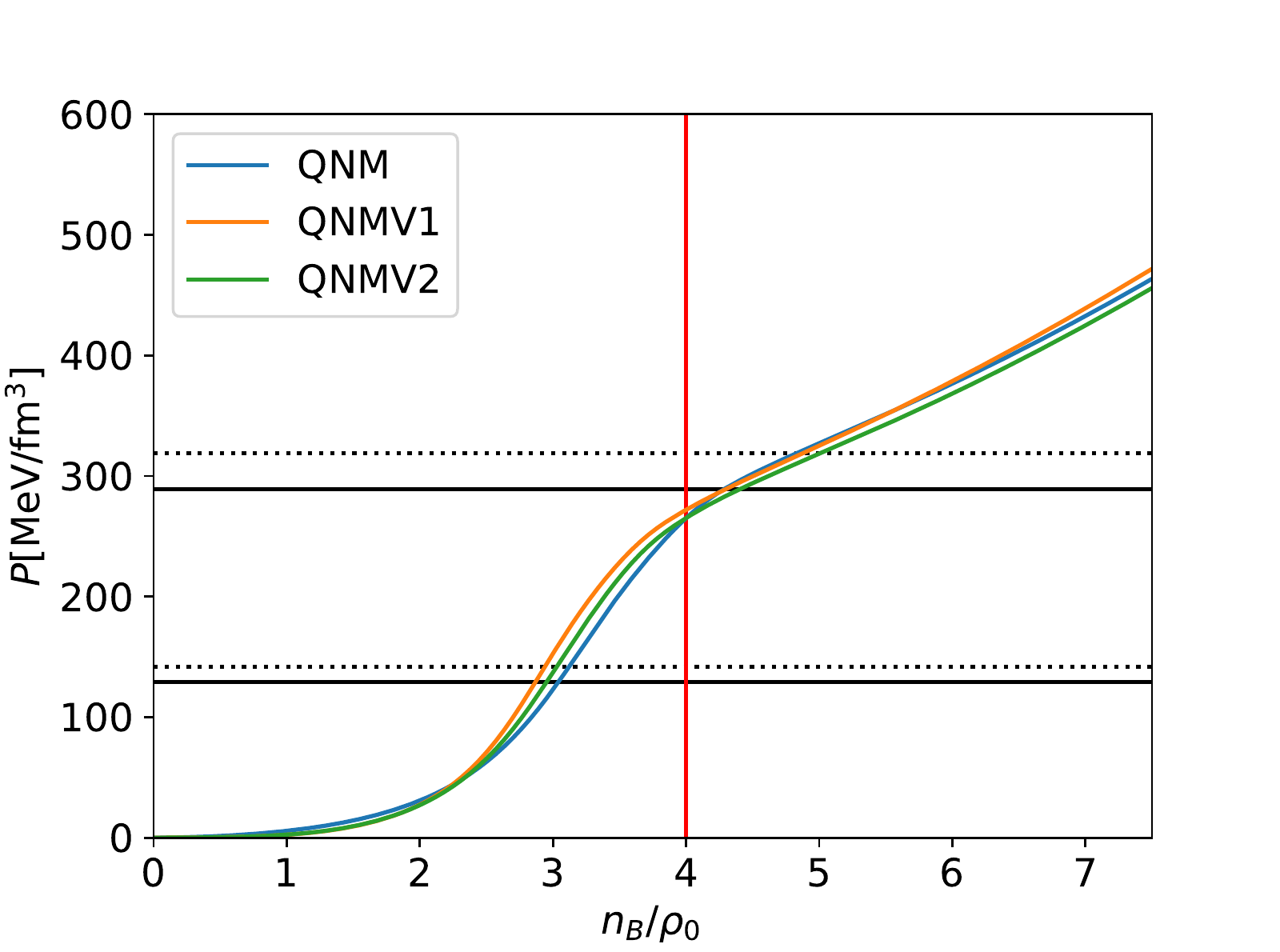}
	\caption{We plot the pressure as a function of baryon density for the three models QNM, QNMV1 and QNMV2. For QNM we use the parameter values $n_0 = 4.5\rho_0$ where $\rho_0$ is the nuclear saturation density and $\Lambda_0 = 0.25$GeV. For QNMV1, we use $n_0=4.0\rho_0$ and $\Lambda_0=0.25$GeV. For QNMV2 we use, $n_0=4.0\rho_0$, $\Lambda_0=0.3$GeV and $\gamma=2$. Also we choose $\tilde{a}=-28.6$ MeV and $\tilde{b}=10$ MeV for QNMV1 and $\tilde{a}=-27.6$ MeV and $\tilde{b}=7.9$ MeV for QNMV2. The red vertical line represents the density $n_B = 4\rho_0$, while the upper and lower bounds on the pressure at $n_B = 4\rho_0$ excluding (including) the NICER measurement of PSR J0030+0451 is shown by horizontal black solid (dashed) lines.}%, and the solid black line indicates the maximum allowed pressure inside neutron stars.}%Some examples of the pressure as a function of baryon density. Legends are as in figure \ref{Pressure around saturation density}. Dashed horizontal lines shows the upper and lower exprimental bounds on the pressure at 4 times the saturation density, and solid horizontal line shows the experimental constraint on the maximum pressure in neutron stars, as proposed in \cite{Stringent_constraints}. }
	\label{pressure}
\end{figure}
\begin{figure}
	\centering
	\includegraphics[width=0.5\textwidth]{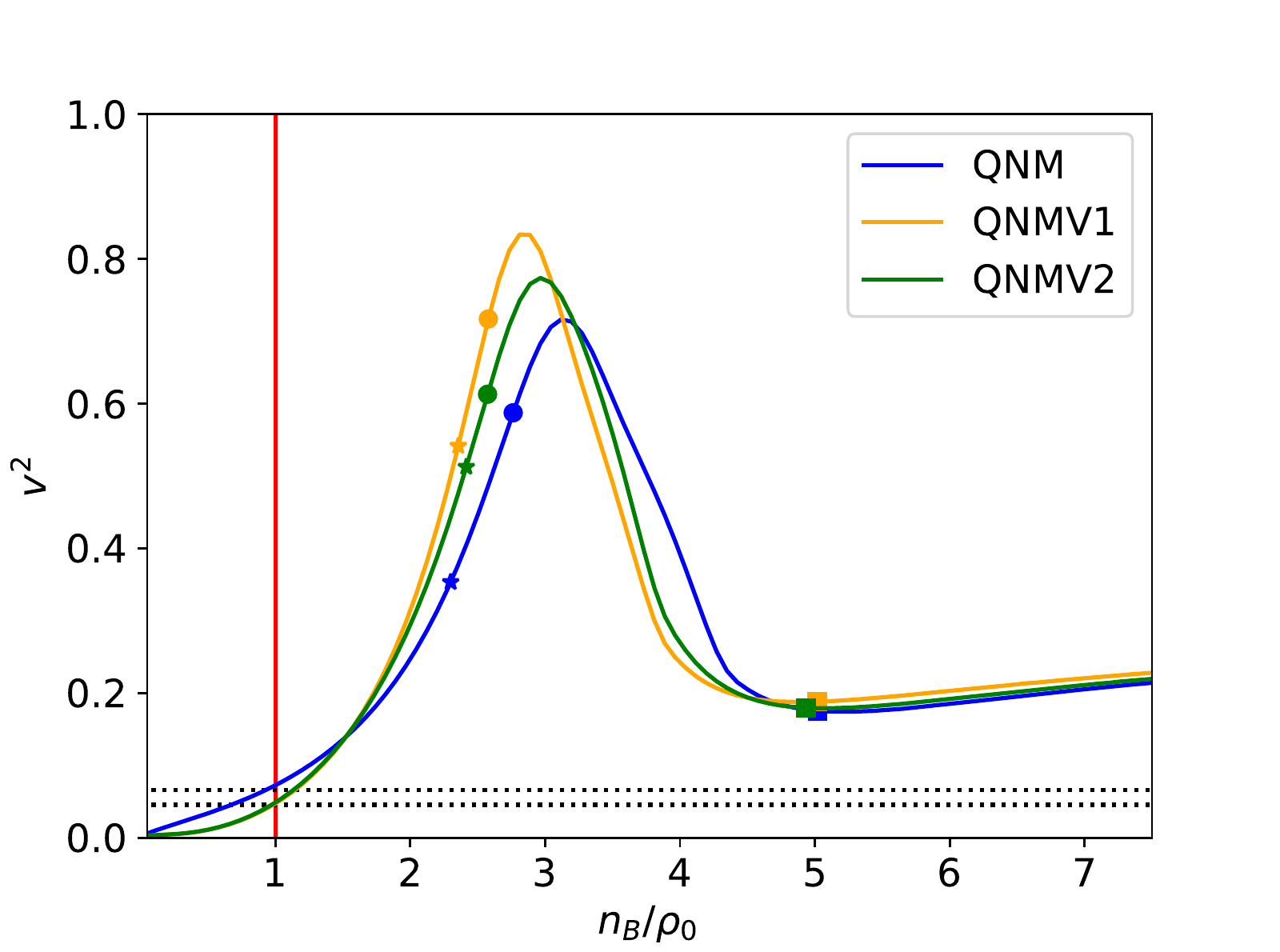}
	\caption{This is a plot showing the speed of sound as a function of baryon density for the three models QNM, QNMV1 and QNMV2. For QNM we use the parameter values $n_0 = 4.5\rho_0$ where $\rho_0$ is the nuclear saturation density and $\Lambda_0 = 0.25$GeV. For QNMV1, we use $n_0=4\rho_0$ and $\Lambda_0=0.25$GeV. For QNMV2 we use, $n_0=4\rho_0$, $\Lambda_0=0.3$GeV and $\gamma=2$. We choose $\tilde{a}=-28.6$ MeV and $\tilde{b}=10$ MeV for QNMV1 and $\tilde{a}=-27.6$ MeV and $\tilde{b}=7.9$ MeV for QNMV2. The red vertical line represents the saturation density, while the horizontal dashed lines show 68\% bands on the speed of sound at the saturation density obtained from many body chiral effective field theory \cite{PhysRevLett.125.202702}. We have indicated the quark onset density as defined in the text by filled stars, the central density of neutron stars with mass 1.4 solar masses by filled circles and the central density of maximum mass stars by filled squares.}
		%Some examples of the speed of sound as a function of baryon density. Legends are as figure \ref{Pressure around saturation density}.}
	\label{v2}
\end{figure}
\begin{figure}
	\centering
	\includegraphics[width=0.5\textwidth]{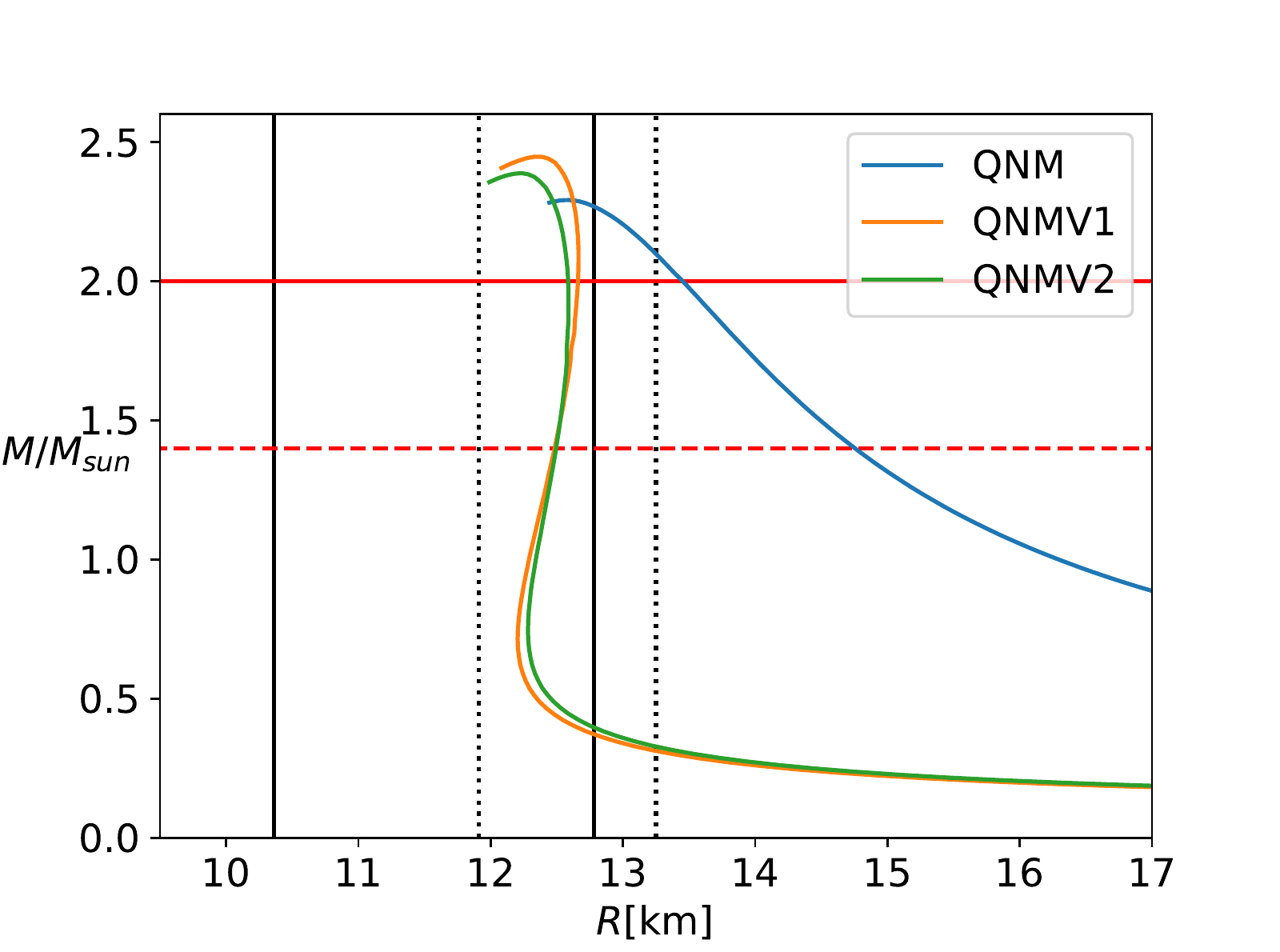}
	\caption{Here we plot the mass-radius relations for the four models QNM, QNMV1 and QNMV2. Neutron stars with $1.4$ ($2$) solar masses lie on the dashed (solid) red line. The black solid(dashed) lines shows the upper and lower bounds on the radii of neutron stars with mass equal to $1.4$ solar masses excluding (including) the NICER measurement of PSR J0030+0451. For QNM we use the parameter values $n_0 = 4.5\rho_0$ where $\rho_0$ is the nuclear saturation density and $\Lambda_0 = 0.25$GeV. For QNMV1, we use $n_0=4\rho_0$ and $\Lambda_0=0.25$GeV. For QNMV2 we use, $n_0=4\rho_0$, $\Lambda_0=0.3$GeV and $\gamma=2$. We choose $\tilde{a}=-28.6$ MeV and $\tilde{b}=10$ MeV for QNMV1 and $\tilde{a}=-27.6$ MeV and $\tilde{b}=7.9$ MeV for QNMV2}
	\label{Mass radius relations}
		%Some examples of mass-radius relations. Dashed lines show the constraint on the radius for neutron stars with mass 1.4 solar masses, while the solid line shows the minimum upper mass limit of 2 solar masses. Legends are as in figure \ref{Pressure around saturation density}.}
\end{figure}

\begin{figure}
	\centering
	\includegraphics[width=0.5\textwidth]{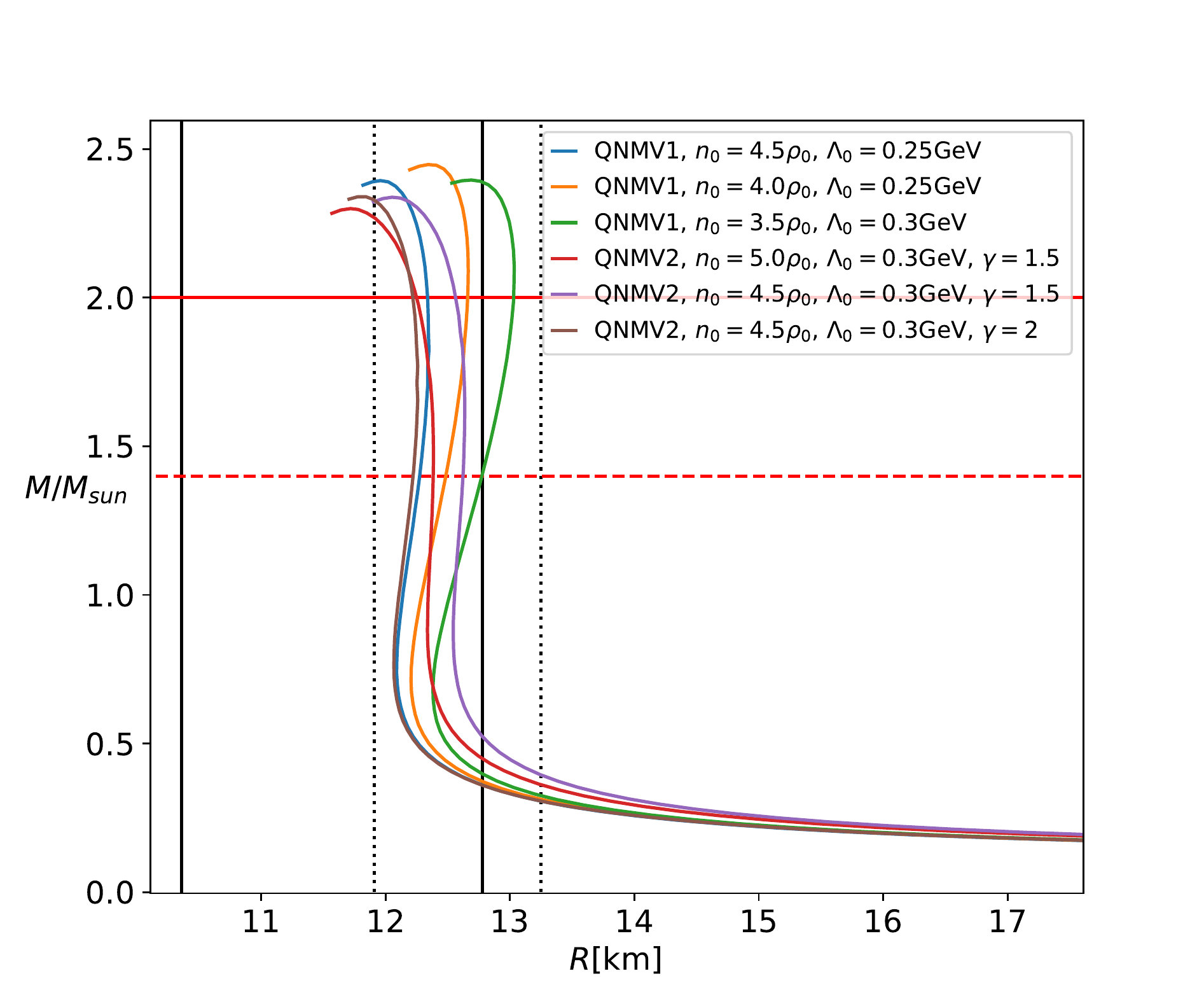}
	\caption{Mass-radius relations for the models QNMV1 and QNMV2 that satisfy all constraints for a collection of choices for $n_0$, $\Lambda_0$ and $\gamma$. Neutron stars with $1.4$ ($2$) solar masses lie on the dashed (solid) red line. The black solid(dashed) lines shows the upper and lower bounds on the radii of neutron stars with mass equal to $1.4$ solar masses excluding (including) the NICER measurement of PSR J0030+0451}
	\label{mass radius all}
\end{figure}
\section{Results and discussion}
From the discussions in the section \ref{nmqm} we know that the models QNMV1 and QNMV2 will satisfy the constraints of pressure and symmetry energy at the saturation density by construction. However, we also need to examine whether the rest of the constraints described in section \ref{3} are satisfied by these models. These constraints include, the radii of neutron stars with masses around 1.4 solar masses which is constrained to be $R_{1.4} = 11.4^{+1.38}_{-1.04}$km ($R_{1.4}^\text{NICER} = 12.54^{+0.71}_{-0.63}$km), the pressure at two and four times the saturation density constrained to be $P_{2\text{sat}} = 14.2^{+18.1}_{-8.4}$ ($P_{2\text{sat}}^\text{NICER} = 28.7^{+15.3}_{-15.0}$MeV/fm$^3$) and  $P_{4\text{sat}}=202^{+87}_{-73}$MeV/fm$^3$ ($P_{4\text{sat}}=211^{+108}_{-69}$MeV/fm$^3$) \cite{Essick:2020flb} and the lower limit on the maximum mass of neutron stars imposed by various observations of neutron stars with $\sim$ $2$ solar masses \cite{Cromartie2020,Antoniadis1233232}. We can now check to what extent the three models, QNM , QNMV1 and QNMV2 as defined in Eq. \ref{q4}, \ref{NMV1} and \ref{e1} respectively, satisfy these constraints. We will also examine the speed of sound for these models as we proceed.
%Having obtained the equations of state that satisfy the constraints of pressure and symmetry energy, we can now check the validity of the equation of state of the different excluded volume models discussed in previous sections. Let's denote these models by the following names for convenience. The excluded volume model of [Larry, KieSang, Sen] which describes symmetric nuclear matter with excluded volume interaction between the nucleons is denoted as SMQM (symmetric matter quarkyonic model). The excluded volume model introduced in this paper where nuclear degrees of freedom are just neutrons as given in is denoted as NMQM(neutron matter quarkyonic model). Similarly, the model where we add the nuclear potential of [Ref] to the NMQM is denoted as NMQMV. And finally, the model of NMQM with a modified excluded volume interaction, i.e. $\gamma>1$ that also includes the nuclear potential is denoted as NMQMV2.
In Figure \ref{Pressure around saturation density} we plot the pressure as a function of the total baryon density $n_B$ normalized by the saturation density $\rho_0$. In order for the equation of state to be within the bounds of experimental constraints obtained from neutron star measurements \cite{PhysRevC.86.015803}, the pressure at the saturation density $P_{\text{sat}}$ should lie within $1.9$ and $2.9$ $\text{MeV}/\text{fm}^3$, as indicated by dashed lines in the figure. In line with expectations, we find that the model QNM (as defined in Eq. \ref{q4}) cannot satisfy the constraints on pressure and symmetry energy at saturation density for any value of $n_0$. 
%for $n_0< 8\rho_0$. If we choose $n_0\sim 8\rho_0$ on the other hand, the pressure at higher baryon densities softens too much to be compatible with the constraint on $P_{4\text{sat}}$. 
In Figure \ref{Pressure around saturation density} we have used the parameters $n_0 = 4.5\rho_0$ and $\Lambda_0 = 0.35$GeV for the model QNM. These parameters are chosen such that QNM satisfies the constraint on $P_{4\text{sat}}$, as shown in figure \ref{pressure}.  %As mentioned earlier, the somewhat high pressure at saturation for the models QSM and NMQM is not necessarily inconsistent with constraints from heavy ion collisions as discussed in [cite 'Constraints on the symmetry energy and neutron skins from experiments and theory']. However, as seen from figure }\ref{Mass radius relations} \hl{the models exhibit rather large radii that are not consistent with constraints on $R_{1.4}$. It is found that lower radii tend to favor lower pressure at saturation, and it has not succeed to find parameters compatible with all the constraints for the models QSM and NMQM}.

In figures \ref{Pressure around saturation density}-\ref{Mass radius relations} we show that QNMV1 and QNMV2 can fit all the constraints discussed in this paper for reasonable parameter values. For QNMV1 we are able to fit all the constraints discussed in the text by fixing $n_0 = 4\rho_0$ and $\Lambda_0 = 0.4$GeV. In the same manner, we are able to fit all constraints for QNMV2 by setting $n_0 = 4\rho_0$, $\gamma = 2$ and $\Lambda_0 = 0.4$GeV. In the plots we have for QNMV1 we use $\tilde{a} = -28.6$MeV and $\tilde{b} = 10.0$MeV for the nucleon potential, while for QNMV2 we have chosen $\tilde{a} = -27.6$MeV and $\tilde{b} = 7.9$MeV in order to keep the symmetry energy for both models around 32MeV. 

In figure \ref{v2} for the speed of sound we indicate the quark onset density, the central density of neutron stars with 1.4 solar masses and the central density of the maximum mass neutron stars. We have defined the quark onset density to be the density at which 1\% of the baryon density is made up of quarks.
	
%We have also included 68\% bands on the speed of sound at two times the saturation density as calculated by} \cite{PhysRevD.79.124032} \hl{in many body perturbation theory at third (fourth) order in the chiral effective field theory expansion, usually referred to as N$^2$LO (N$^3$LO). Neither of the models seems to lie within these bounds, but it should be noted the authors suggest that these bounds might be on the low end due to edge effects that will go away as the machine learning algorithm is trained on data at higher densities. Nevertheless, if one insists on satisfying these bounds, it is found that QNMV1 can at least do so for the N$^2$LO bounds if we remove the 9 first terms in the expansion of Eq.} \ref{q4a}. \hl{This suggests that a possible discrepancy between our models and }\cite{PhysRevD.79.124032} \hl{can be removed by modifying the nuclear potential. We will leave further discussion of this for future work.}

For the mass-radius plots we have taken the equations of state of the inner and outer crust to be given by \cite{2001A&A...380..151D} and \cite{1971ApJ...170..299B}, respectively, and we have patched them together with our equation of state for the core in a similar manner as in \cite{PhysRevD.79.124032}. %\hl{It should be mentioned that it has been pointed out by} \cite{2016PhRvC..94c5804F} \hl{that the choice of patching procedure can make a difference of up to 1km in the radii of neutron stars with a mass of 1 solar masses. However, we will not further analyze the details of the patching procedure since we consider it to be beyond the scope of this text.} 
We note that $P_{0}$ and the radius at 1.4 solar masses $R_{1.4}$ are very sensitive to $n_0$ and $\gamma$. It is found that both $P_{0}$ and $R_{1.4}$ decrease with increasing values of $n_0$ and $\gamma$. In addition, the pressure at densities beyond the saturation density, the maximum of the speed sound, as well as the maximum mass decrease with increasing values of $\Lambda_0$. 

Since we have fitted the same equation of state for the crust for all the three models, one would expect that the mass-radius curves of QNM, QNMV1 and QNMV2 approach one another for neutron stars with low central density. However, it is clear from figure \ref{Mass radius relations} that QNM is very different from QNMV1 and QNMV2 even in this regime. The reason for this is that the equation of state for QNM is so stiff at low densities that in order to keep the pressure continuous as a function of baryon density, the crust-core matching density becomes around $0.01\rho_0$, roughly an order of magnitude smaller than it is for QNMV1 and QNMV2. Note that this is expected since the QNM model does not include the appropriate nuclear interactions at low density. This of course is tied to QNM model violating the mass radius constraints coming from NS observations. We present the QNM mass-radius plot in figure \ref{Mass radius relations} in order to highlight the importance of the approaches adopted in the construction of QNMV1 and QNMV2 models.

We finish this section by showing mass-radius curves for a range of different parameter values for QNMV1 and QNMV2 that fit the neutron star constraints in Figure \ref{mass radius all}. 
%We note that the most stringent requirement for the models QNMV1 and QNMV2 is the radii for neutron stars with 1.4 solar masses. In order for the radii to be small enough, we need the pressure around saturation to be on the lower end of the allowed values. 
%We note that this may be an issue due to the apparent discrepancy in constraints on $P_{\text{sat}}$ between neutron star observations and heavy ion collisions. 
\section{Conclusion:}
In this paper we set out to explore the effectiveness of the excluded volume model proposed in \cite{PhysRevC.101.035201} in describing the mass radius relations of neutron stars. Although there have been other models of quarkyonic matter which have been successful in describing several neutron star properties \cite{Han:2019bub, Zhao:2020dvu}, the model of \cite{PhysRevC.101.035201} is one of the few available dynamical models of quarkyonic matter. For other examples of dynamical model for quarkyonic matter, see \cite{Kovensky:2020xif, Cao:2020byn}. More specifically, the shell width of quarkyonic matter and its variation arise dynamically in the model of \cite{PhysRevC.101.035201}. This is an extremely attractive feature of the excluded volume dynamical model since we do expect the shell width to be a function of the parameters of the theory including the baryon density. This feature also makes this model particularly suitable for incorporating finite temperature corrections as shown in \cite{Sen:2020peq}. It is expected, as found in this paper, that the simple model of \cite{PhysRevC.101.035201} will not describe the detailed properties of neutron stars unless appropriate changes are incorporated in the model. 
We found in this paper that the model in \cite{PhysRevC.101.035201} when augmented with the right nuclear interactions at low density can appropriately fit the constraints of neutron star mass and radius. However, our model does not take into account beta equilibrium which should be incorporated in the excluded volume dynamical model for completeness. A nondynamical quarkyonic matter model including beta equilibrium was constructed in \cite{Zhao:2020dvu} which satisfied neutron star constraints after conditions of chemical equilibrium and beta equilibrium were imposed. Whether a dynamical excluded volume model with appropriate nuclear interactions at low density can correctly describe beta equilibrium while also respecting neutron star mass radius constraint should be explored in future work. Similarly, to make more realistic models of quarkyonic matter one also needs to incorporate strange quarks as degrees of freedom in the quark sector. Some previous work \cite{Duarte:2020xsp, Duarte:2020kvi} have explored strange quark matter and beta equilibrium within the context of excluded volume quarkyonic matter. However, there are quite a few differences between the approach followed in this paper and in \cite{Duarte:2020xsp, Duarte:2020kvi}. For example, \cite{Duarte:2020xsp, Duarte:2020kvi} incorporate a density independent regulator for the quark density of states which gives rise to quarks at relatively low baryon density. Also, \cite{Duarte:2020kvi} uses the Maxwell construction to describe low density properties of nuclear matter while implementing an interpolation procedure to smoothly connect low density and high density regimes of quarkyonic matter. In this work, we don't have to use any interpolation procedure: low density nuclear interactions are captured within the excluded volume model of quarkyonic matter itself. We believe the models constructed in this paper are especially suitable for finite temperature analysis of quarkyonic matter. We plan to incorporate finite temperature corrections in the equations of state presented here so that the behavior of the speed of sound as a function of temperature can be examined. 
\section{Acknowledgment}
We thank Sanjay Reddy for comments on the manuscript. This work was supported by Iowa State University Startup funds.

\bibliography{bibliography2}
\end{document}